%% file: yaglikca2021blockhammer_arxiv.tex
\documentclass[camera,letterpaper,nomarginnotes,nonarrowgutter]{jpaper}
\usepackage{mathptmx} %
\usepackage{courier}
\usepackage[italic]{mathastext}
\usepackage[dvipsnames]{xcolor}
\usepackage{fancyhdr}
\usepackage[normalem]{ulem}
\usepackage[sort]{cite}
\usepackage{siunitx}
\usepackage{tikz}
\usepackage{multirow}
\usepackage{enumitem}
\usepackage{algorithm}
\usepackage{listings}
\usepackage{amsmath}
\usepackage{todonotes}
\usepackage{arydshln} %
\usepackage{xspace}
\usepackage{soul} %
\usepackage[bookmarks=true,breaklinks=true,hidelinks]{hyperref}
\usepackage{amssymb}%
\usepackage{pifont}%
\usepackage{setspace}
\usepackage{footmisc}
\usepackage{balance}
\usepackage{hhline} %
\usepackage[all]{nowidow}
\usepackage[compact]{titlesec}
\usepackage{footnote}

\titlespacing\section{0pt}{2pt plus 0.5pt minus 0.5pt}{-1pt plus 0.5pt minus 0pt}
\titlespacing\subsection{0pt}{2pt plus 0.5pt minus 1pt}{0pt plus 0.5pt minus 0pt}
\titlespacing\subsubsection{0pt}{2pt plus 0.5pt minus 1pt}{2pt plus 0.5pt minus 1pt}

\makeatletter
\g@addto@macro{\normalsize}{%
  \setlength{\abovedisplayskip}{2pt plus 1pt minus 1pt}
  \setlength{\belowdisplayskip}{2pt plus 1pt minus 1pt}
  \setlength{\abovedisplayshortskip}{0pt}
  \setlength{\belowdisplayshortskip}{0pt}
  \setlength{\intextsep}{2pt plus 1pt minus 1pt}
  \setlength{\textfloatsep}{3pt plus 1pt minus 1pt}
  \setlength{\dbltextfloatsep}{3pt plus 1pt minus 1pt}
  \setlength{\skip\footins}{4pt plus 1pt minus 1pt}}
\makeatother

\renewcommand\thefootnote{{\arabic{footnote}}}

\definecolor{darkwarmgray}{rgb}{0.15, 0.050, 0.05}
\newcommand{\circled}[1]{{\tikz[baseline=(char.base)]{\node[shape=circle,inner sep=1pt,fill=darkwarmgray, text=white] (char) {\footnotesize \textbf{#1}};}}}

\newcommand{\cmark}{\ding{51}}
\newcommand{\xmark}{\ding{55}}

\newcommand{\figref}[1]{Figure~\ref{fig:#1}}
\newcommand{\secref}[1]{Section~\ref{sec:#1}}

\newcommand{\subsubsecref}[1]{Section~\ref{subsubsec:#1}}
\newcommand{\equref}[1]{Equation~\ref{equ:#1}}
\newcommand{\tabhead}[2][10em]{%
  \rotatebox{90}{\parbox{#1}{\raggedright \textbf{#2}}}}
\newcommand{\thead}[1]{\tabhead[20mm]{#1}}

\newcommand{\rbbl}{{RowBlocker-BL}}
\newcommand{\rbhb}{{RowBlocker-HB}}
\newcommand{\prohit}{{PRoHIT}}
\newcommand{\para}{{PARA}}
\newcommand{\cbt}{{CBT}}
\newcommand{\twice}{{TWiCe}}
\newcommand{\graphene}{{Graphene}}
\newcommand{\mrloc}{{MRLoc}}

\newcommand{\trc}{$t_{RC}$}
\newcommand{\tfaw}{$t_{FAW}$}
\newcommand{\trefw}{$t_{REFW}$}

\newcommand{\nbl}{$N_{BL}$}
\newcommand{\nep}{$\nepn{}$}

\newcommand{\nrh}{$N_{RH}$}

\newcommand{\tdelay}{$\tdelayn{}$}
\newcommand{\rhli}{$RHLI$}
\newcommand{\nepn}{N_{ep}}

\newcommand{\trefwn}{t_{REFW}}

\newcommand{\nthn}{\nbln{}}
\newcommand{\nbln}{N_{BL}}
\newcommand{\nrhn}{N_{RH}}
\newcommand{\trcn}{t_{RC}}

\newcommand{\tfawn}{t_{FAW}}

\newcommand{\tbfn}{t_{CBF}}
\newcommand{\tbf}{$\tbfn{}$}
\newcommand{\cbfan}{CBF_{A}}
\newcommand{\cbfa}{$\cbfan$}
\newcommand{\cbfbn}{CBF_{B}}
\newcommand{\cbfb}{$\cbfbn$}

\newcommand{\tdelayn}{t_{Delay}}

\newcommand{\tepochn}{t_{ep}}

\newcommand{\nrhtuned}{$\nrhtunedn{}$}
\newcommand{\nrhtunedn}{{N_{RH}}^*}

\newcommand{\naggn}{N_{ep}}
\newcommand{\nepmaxin}{N_{ep_{max}}}

\newcommand{\nepmax}{$\nepmaxin{}$}
\newcommand{\npren}{N_{ep-1}}

\newcommand{\nblpn}{{N_{BL}}^{*}}

\newcommand{\nagg}{$\naggn{}$}
\newcommand{\npre}{$\npren{}$}

\newcommand{\nblp}{$\nblpn{}$}
\usepackage{dblfloatfix}

\def\UrlBreaks{\do\/\do-\/\do.\/\do:}

\expandafter\def\expandafter\UrlBreaks\expandafter{\UrlBreaks
  \do\a\do\b\do\c\do\d\do\e\do\f\do\g\do\h\do\i\do\j
  \do\k\do\l\do\m\do\n\do\o\do\p\do\q\do\r\do\s\do\t
  \do\u\do\v\do\w\do\x\do\y\do\z\do\A\do\B\do\C\do\D
  \do\E\do\F\do\G\do\H\do\I\do\J\do\K\do\L\do\M\do\N
  \do\O\do\P\do\Q\do\R\do\S\do\T\do\U\do\V\do\W\do\X
  \do\Y\do\Z}

\newcommand{\squishlist}{
 \begin{list}{$\circ$}
  { \setlength{\itemsep}{0pt}
     \setlength{\parsep}{0pt}
     \setlength{\topsep}{0pt}
     \setlength{\partopsep}{0pt}
     \setlength{\leftmargin}{1em}
     \setlength{\labelwidth}{1em}
     \setlength{\labelsep}{0.5em} } }

\newcommand{\squishsublist}{
\begin{list}{$\rightarrow$}
 { \setlength{\itemsep}{0pt}
    \setlength{\parsep}{0pt}
    \setlength{\topsep}{-10em}
    \setlength{\partopsep}{-3pt}
    \setlength{\leftmargin}{1em}
    \setlength{\labelwidth}{1em}
    \setlength{\labelsep}{0.5em} } }

\newcommand{\squishend}{
  \end{list}  }

\fancypagestyle{firstpage}{

  \fancyhead[C]{\normalsize{Extended Version of HPCA 2021 Paper -- Confidential Draft -- Do NOT Distribute!!\versiontext{}}\vspace{-3em}}
  \pagenumbering{arabic}
}

\newcommand{\affilETH}[0]{\small {$^1$}}
\newcommand{\affilUIUC}[0]{{\small {$^2$}}}
\title{BlockHammer: Preventing RowHammer at Low Cost\\\vspace{-0.3em}by Blacklisting Rapidly-Accessed DRAM Rows}
\author{\vspace{-17pt}\\%
\fontsize{11}{12}\selectfont%
{A. Giray Ya\u{g}l{\i}k\c{c}{\i}\affilETH{}}\quad%
{Minesh Patel\affilETH{}}\quad%
{Jeremie S. Kim\affilETH{}}\quad%
{Roknoddin Azizi\affilETH{}}\quad%
{Ataberk Olgun\affilETH{}}\quad%
{Lois Orosa\affilETH{}}%
\vspace{-1pt}\\%
\fontsize{11}{12}\selectfont%
{Hasan Hassan\affilETH{}}\quad%
{Jisung Park\affilETH{}}\quad%
{Konstantinos Kanellopoulos\affilETH{}}\quad%
{Taha Shahroodi\affilETH{}}\quad%
{Saugata Ghose\affilUIUC{}}\quad%
{Onur Mutlu\affilETH{}}%
\vspace{-1pt}\\%
{\fontsize{10}{11}\selectfont
\affilETH\emph{ETH Z{\"u}rich}%
\qquad\quad%
\affilUIUC\emph{University of Illinois at Urbana--Champaign}%
}
\vspace{-16pt}\vspace{0.3em}}

\begin{document}
\bstctlcite{IEEEexample:BSTcontrol}
\maketitle

\thispagestyle{plain}
\pagestyle{plain}

\setstretch{0.77}
\input{00_abstract}
\input{01_introduction}
\setstretch{0.77}
\input{02_background}
\setstretch{0.78}
\input{03_blockhammer}
\input{04_many_sided_attacks}
\setstretch{0.775}
\input{05_security_analysis}
\setstretch{0.77}
\input{06_areaandlatency}
\setstretch{0.77}
\input{07_methodology}
\setstretch{0.774}
\input{08_experimental_evaluation}
\setstretch{0.775}
\input{09_qualitative_analysis}
\setstretch{0.77}
\input{10_relatedwork}
\input{11_conclusion}
\section*{Acknowledgments}
We thank the anonymous reviewers of HPCA 2020, ISCA 2020, MICRO 2020, and HPCA 2021 for {feedback}. We thank the
SAFARI Research Group members for {valuable} feedback and the stimulating intellectual environment they provide. We acknowledge the generous gifts provided by our industrial partners: Google, Huawei, Intel, Microsoft, and VMware.

\balance
\setstretch{0.8}
\bibliographystyle{IEEEtranS}
\bibliography{yaglikca2021blockhammer_arxiv}
\pagebreak
\nobalance
\onecolumn
\setstretch{1.0}
\appendix
\input{12_appendix_tables}

\end{document}

%% file: 00_abstract.tex
\begin{abstract}
  Aggressive memory density scaling causes modern DRAM devices to suffer from RowHammer, a phenomenon where rapidly activating (i.e., hammering) a DRAM row can cause bit-flips in physically-nearby rows.
  Recent studies demonstrate that modern DDR4/LPDDR4 DRAM chips, including chips previously marketed as RowHammer-safe, are even more vulnerable to RowHammer than older DDR3 DRAM chips.
  Many works show that attackers can exploit RowHammer bit-flips to reliably mount system-level attacks to escalate privilege and leak private data.
  Therefore, it is critical to ensure RowHammer-safe operation on all DRAM-based systems as they become increasingly more vulnerable to RowHammer.
  Unfortunately, state-of-the-art RowHammer mitigation mechanisms face two major challenges.
  First, they incur increasingly higher performance and/or area overheads when applied to more vulnerable DRAM chips.
  Second, they require either closely-guarded proprietary information about the DRAM chips' physical circuit layouts or modifications to the DRAM chip design.

  In this paper, we show that it is possible to efficiently and scalably prevent RowHammer bit-flips without knowledge of or modification to DRAM internals.
  To this end, we introduce BlockHammer, a low-cost, effective, and easy-to-adopt RowHammer mitigation mechanism that prevents all RowHammer bit-flips while overcoming the two key challenges.
  BlockHammer selectively throttles memory accesses that could otherwise potentially cause RowHammer bit-flips.
  The key idea of BlockHammer is to (1) track row activation rates using area-efficient Bloom filters, and (2) use the tracking data to ensure that no row is ever activated rapidly enough to induce RowHammer bit-flips.
  By guaranteeing that no DRAM row ever experiences a RowHammer-unsafe activation rate, BlockHammer (1) makes it impossible for a RowHammer bit-flip to occur and (2) greatly reduces a RowHammer attack's impact on the performance of co-running benign applications.
  Our evaluations across a comprehensive range of 280~workloads show that, compared to the best of six state-of-the-art RowHammer mitigation mechanisms (all of which require knowledge of or modification to DRAM internals), BlockHammer provides
  (1)~competitive performance and energy when the system is not under a RowHammer attack and
  (2)~significantly better performance and energy when the system is under a RowHammer attack.
  BlockHammer's source code is openly and freely available~\cite{blockhammergithub}.
\end{abstract}

%% file: 01_introduction.tex
\section{Introduction}
\label{sec:introduction}

Improvements to manufacturing process technology {have} increased DRAM storage density by reducing DRAM cell size and cell-to-cell spacing for decades.
Although such optimizations improve a DRAM chip's cost-per-bit, they negatively impact DRAM reliability~\cite{mutlu2013memory, meza2015revisiting}. 
Kim et al.~\cite{kim2014flipping} show that modern DRAM chips are susceptible to the RowHammer phenomenon, where opening and closing (i.e., \emph{activating} and \emph{precharging}) a DRAM row (i.e., \emph{aggressor row}) at a \emph{high enough} rate (i.e., \emph{hammering}) can cause bit-flips in physically-nearby rows (i.e., {\emph{victim rows}})~\cite{redeker2002investigation, mutlu2017rowhammer, yang2019trap, mutlu2019rowhammer}.
Many works demonstrate various system-level attacks using RowHammer to escalate privilege or {leak} private data (e.g.,~\cite{seaborn2015exploiting, van2016drammer, gruss2016rowhammer, razavi2016flip, pessl2016drama, xiao2016one, bosman2016dedup, bhattacharya2016curious, qiao2016new, jang2017sgx, aga2017good,
mutlu2017rowhammer, tatar2018defeating ,gruss2018another, lipp2018nethammer,
van2018guardion, frigo2018grand, cojocar2019eccploit,  ji2019pinpoint, mutlu2019rowhammer, hong2019terminal, kwong2020rambleed, frigo2020trrespass, cojocar2020rowhammer, weissman2020jackhammer, zhang2020pthammer, rowhammergithub, yao2020deephammer}).
{Recent} findings {indicate} that RowHammer is a more serious problem than ever and {that} it is {expected to worsen for future DRAM chips}~\cite{mutlu2017rowhammer, mutlu2019rowhammer, kim2020revisiting}.
{Therefore,} comprehensively protecting {DRAM} against all types of RowHammer attacks is essential for {the security and reliability of} current and future DRAM-based {computing} systems.

\setstretch{0.8}
{Although DRAM {vendors} {{currently implement}}} {in-DRAM RowHammer mitigation mechanisms, e.g., target row refresh~\cite{lee2014green, micron2014ddr4, jedec2015low, jedec2015hbm, jedec2017, frigo2020trrespass}},
recent works report that commodity {DDR3~\cite{park2016statistical}, DDR4~\cite{pessl2016drama, aga2017good, frigo2020trrespass, cojocar2020rowhammer, kim2020revisiting}, and LPDDR4~\cite{kim2020revisiting}} chips {remain vulnerable to RowHammer}.
In particular,  TRResspass~\cite{frigo2020trrespass} shows that an attacker can still reliably {induce} RowHammer {bit-flips} {in commodity (LP)DDRx DRAM chips} by circumventing the in-DRAM mitigation mechanisms. Kim et al.~\cite{kim2020revisiting} show that from 2014 to 2020, DRAM chips {have become} significantly more vulnerable to RowHammer {bit-flips}, {with} over an order of magnitude reduction in the required number of {row} activations to induce a {bit-flip} (from 139.2k to 9.6k).

Given the severity of RowHammer, {various} mitigation methods have been proposed, which we classify into four {high-level} approaches:
($i$)~\emph{increased refresh rate}, which {refreshes \emph{all}} rows more frequently to reduce the probability of a successful {bit-flip}~\cite{AppleRefInc, kim2014flipping};
($ii$)~\emph{physical isolation}, which {physically separates sensitive data from {any} potential attacker's memory space (e.g., by adding buffer rows between sensitive data regions and other data)~\cite{konoth2018zebram, van2018guardion, brasser2017can};}
($iii$)~\emph{reactive refresh}, which observes row activations
and refreshes the potential victim rows as a reaction to rapid row activations~\cite{kim2014flipping, aweke2016anvil, son2017making, seyedzadeh2018cbt, you2019mrloc, lee2019twice, park2020graphene}; and
{($iv$)~\emph{proactive throttling}, which {limits} {row activation} rates~\cite{greenfield2012throttling, mutlu2018rowhammer, kim2014flipping} to {RowHammer-safe levels}.}
Unfortunately, each of these four approaches faces at least one of two major challenges towards effectively mitigating RowHammer.

\noindent
\textbf{{Challenge} 1: {Efficient Scaling as RowHammer Worsens.}}
{As DRAM chips become more vulnerable to RowHammer {(i.e., RowHammer {bit-flips} can occur at significantly lower row activation {counts} than before)}, {{mitigation mechanisms} need to act more aggressively.}}
{A} \emph{scalable} mechanism {should} exhibit acceptable performance, energy, and area overheads as its design is reconfigured for more vulnerable DRAM chips. Unfortunately, as chips become more vulnerable to RowHammer, {most} state-of-the-art mechanisms of all four approaches either cannot easily adapt because they are based on fixed design points, or their performance, energy, and/or area overheads become increasingly significant.
($i$)~{Increasing the refresh rate further {in order to prevent all RowHammer {bit-flips}} is
{prohibitively expensive,}
even for {existing} DRAM chips~\cite{kim2020revisiting}, due to the large number of rows that {must} be refreshed within a refresh window.
{($ii$)}~P}hysical isolation mechanisms {must} provide {greater} isolation (i.e., increase the physical distance) between sensitive data and a potential attacker's memory space {as DRAM chips become denser and more vulnerable to RowHammer}.
This is because
denser chip designs bring circuit elements closer together,
which increases the {number of} rows across which the hammering {of} an aggressor row can induce RowHammer {bit-flips}~\cite{kim2014flipping, mutlu2017rowhammer, yang2019trap, kim2020revisiting}. {P}roviding {greater} isolation (e.g., increasing the number of buffer rows between sensitive data and an attacker's memory space) both wastes increasing amounts of memory capacity and reduces the fraction of physical memory that can be protected from RowHammer attacks.
{($iii$)}~Reactive refresh mechanisms need to increase the rate at which they refresh potential victim {rows.} {P}rior {work~\cite{kim2020revisiting}} shows that state-of-the-art {reactive refresh} RowHammer mitigation mechanisms
{lead to prohibitively large performance overheads with increasing RowHammer vulnerability.}
{($iv$)}~{Existing proactive} throttling approaches {must throttle {activations} at a more aggressive rate to counteract the increased RowHammer vulnerability.}
{This requires either throttling} row activations of benign applications {as well} or {tracking} per-row activation rates for the entire refresh window,
{incurring prohibitively-expensive performance or area overheads} {even {for} existing} DRAM chips~\cite{kim2014flipping, mutlu2018rowhammer}.

\setstretch{0.77}
\noindent
\textbf{{Challenge} 2: Compatibility with Commodity DRAM Chips.}
Both {($ii$)} physical isolation and {($iii$)} reactive refresh mechanisms require the ability to either (1)~identify \emph{all potential victim rows} that can be affected by hammering a given row or (2)~modify the DRAM chip
such that {either} the potential victim rows are internally isolate{d} within the {DRAM} chip or the RowHammer mitigation mechanism can accurately issue reactive refreshes to {all} potential victim {rows}.
{Identifying all potential victim {rows} requires knowing the mapping {schemes} that {the DRAM chip} uses to internally {translate} memory-controller-visible row addresses to physical row addresses~\cite{kim2014flipping, smith1981laser, horiguchi1997redundancy, keeth2001dram, itoh2013vlsi, liu2013experimental,seshadri2015gather, khan2016parbor, khan2017detecting, lee2017design, tatar2018defeating, barenghi2018software, cojocar2020rowhammer,  patel2020beer}}.
{Unfortunately,} DRAM vendors consider their in-DRAM row address mapping schemes to be highly \emph{proprietary} and do not reveal any details in {publicly-available} documentation, as these details contain insights into the chip design and manufacturing quality~\cite{smith1981laser, horiguchi1997redundancy, keeth2001dram, itoh2013vlsi, lee2017design, patel2020beer}
(discussed in Section~\ref{sec:background_physicallayout}).
As a result, both physical isolation and reactive {refresh} are limited to systems that can {(1)}~obtain such proprietary information on {in-}DRAM row address mapping or {(2)}~modify DRAM chips internally.

\textbf{Our goal} in this paper is to design a {low-cost, effective,} and easy-to-adopt RowHammer mitigation mechanism that (1)~scales efficiently with worsening RowHammer vulnerability to prevent RowHammer {bit-flips} in current and future DRAM chips,
and (2)~is {seamlessly} compatible with \emph{commodity} DRAM chips{,} without requiring {proprietary information about or} modifications to {DRAM chips}.
To this end, we propose BlockHammer, a new {proactive throttling-based} RowHammer mitigation mechanism, which is openly and freely available~\cite{blockhammergithub}.
{BlockHammer's key idea is} to track row activation rates using area-efficient Bloom filters and use the tracking data to ensure that no row is ever activated rapidly enough to induce RowHammer {bit-flips}.
Because BlockHammer requires no proprietary information about or modifications to DRAM chips, it can be implemented completely within the {memory controller}.
Compared to prior works that require proprietary information or DRAM chip modifications, BlockHammer {provides} (1)~competitive performance and energy when the system is not {under a {RowHammer}} attack and (2)~significantly better performance and energy {({average/maximum} of 45.0\%{/61.9\%} and 28.9\%{/33.8\%}, respectively)} when the system {\emph{is}} {under a RowHammer} attack. To our knowledge, this is the first work that prevents RowHammer {bit-flips} efficiently and scalably without {knowledge} of or modification to DRAM internals.

\noindent
\textbf{Key Mechanism.} {BlockHammer consists of two components: \emph{RowBlocker} and \emph{AttackThrottler}.}
\emph{RowBlocker} {tracks and limits the activation rates of DRAM rows {to a rate lower than
at} which RowHammer bit-flips begin to occur, i.e., the \emph{RowHammer threshold} (\nrh{}).}
{To track activation rates in an area-efficient manner,}
RowBlocker employs {a false-negative-free variant of} counting Bloom filters~\cite{fan2000summary, li2012compression} {that eliminates the need for per-row counters.}
{When RowBlocker observes that a row's activation count} {within a given time interval exceeds a predefined threshold} {(which we set to be smaller than
{\nrh{}})},
{RowBlocker {\emph{blacklists} the row, i.e.,}
{flags the row as a potential {aggressor row} and limits further activations to the row}
until the end of the time interval, {ensuring that} the row's overall activation rate {never reaches} a RowHammer-unsafe level.}
As a result, RowBlocker ensures that a successful {RowHammer} attack is impossible.

{\emph{AttackThrottler} alleviates the performance degradation {a RowHammer attack {imposes on} benign applications.}
To do so, AttackThrottler reduces the memory bandwidth usage of an attacker thread by applying a quota to the thread's total number of in-flight memory requests for {a determined} time {period}. AttackThrottler {sets} the quota for each thread inversely proportional to the rate at which the thread activates a blacklisted row. As a result, AttackThrottler reduces the memory bandwidth consumed by an attacker, thereby allowing concurrently-running benign applications to have {higher {performance}} when accessing memory.} {To further mitigate the performance impact of RowHammer attacks, AttackThrottler can optionally} expose {the rate at which each thread activates a blacklisted row}
to the operating system {(OS)}. {{This information can} be used as a {dependable} indicator of a thread's likelihood of performing a RowHammer attack, enabling the OS to} employ more sophisticated {thread} scheduling and quality-of-service support.

We evaluate BlockHammer's (1)~security guarantees via a mathematical proof in \secref{mech_security}; (2)~area, {{static} power, access energy,} and latency overheads for storing and accessing metadata by using circuit models~\cite{muralimanohar2009cacti6, synopsys} in \secref{blockhammer_areaoverhead}; {and} (3)~performance and {DRAM} energy {overheads} using {cycle-level}
simulations~\cite{Kim2016Ramulator, ramulatorgithub, drampower} {in \secref{evaluation}}.
{Our evaluations {for a realistic RowHammer threshold (32K activations {within a \SI{64}{\milli\second} refresh window}~\cite{kim2020revisiting})} show that BlockHammer guarantees RowHammer-safe operation with {only} 0.06\% area, {0.7\% performance}, and {0.6\%} {DRAM} energy overheads for benign (i.e., non-attacking) workloads,
compared to a baseline system with no RowHammer mitigation}.
When a RowHammer attack {exists} within a multiprogrammed workload, BlockHammer successfully identifies and throttles {the} attacker's {row activations} with 99.98\% accuracy,
{resulting} in a 45.0\% {average}
improvement in the performance of concurrently-running benign applications.
{We show that}
BlockHammer {more efficiently scales} {with increasing RowHammer vulnerability} {than six state-of-the-art RowHammer mitigation mechanisms,} {without requiring knowledge of or {modification} to the internals}
of DRAM chips.

{{Building} on analyses done by prior work on RowHammer mitigation~\cite{kim2014flipping, gruss2018another, mutlu2017rowhammer, mutlu2018rowhammer, mutlu2019rowhammer, kim2020revisiting}, we {describe} in \secref{qualitative_analysis} that a low-cost, effective, and easy-to-adopt RowHammer mitigation mechanism must:}
(1)~{address} a \emph{comprehensive threat model}, (2)~{be {seamlessly} compatible} with {\emph{commodity}} DRAM chips {(i.e., {require no knowledge of or modifications to} DRAM chip internals)}, (3)~\emph{scale} {efficiently} with increasing RowHammer vulnerability, {and} (4)~\emph{deterministically} {prevent} all RowHammer attacks.
We find that, among all {14}
RowHammer mitigation mechanisms that we examine, BlockHammer is the \emph{only} {one} that satisfies all {four} key properties.

We make the following contributions in this work:
\squishlist
 \item{We introduce the first mechanism that efficiently and scalably prevents RowHammer bit-flips \emph{without} knowledge of or modification to DRAM internals.
 Our mechanism, BlockHammer, provides competitive performance and energy with existing RowHammer mitigation mechanisms when the system is \emph{not} under a RowHammer attack, and \emph{significantly} better performance and energy than existing mechanisms when the system \emph{is} under a RowHammer attack.}
 \item {We show that a proactive throttling approach to prevent RowHammer bit-flips can be implemented efficiently using Bloom filters. We employ a variant of counting Bloom filters that (1)~avoids the area and energy overheads of per-row counters used by prior proactive throttling mechanisms, and (2)~never fails to detect a RowHammer attack.}
 \item {We show that we can greatly reduce the performance degradation and energy wastage a RowHammer attack inflicts on benign threads and the system by accurately identifying the RowHammer attack thread and reducing its memory bandwidth usage. We introduce a new metric called the \emph{RowHammer likelihood index}, which enables the memory controller to distinguish a RowHammer attack from a benign thread.}
\squishend

%% file: 02_background.tex
\section{Background}

This section provides a concise overview of (1)~DRAM organization and operation, (2)~the RowHammer phenomenon, and (3)~in-DRAM row address {mapping}. {{{For more detail, we} refer the reader to p}rior works {on DRAM and RowHammer}~\cite{liu2012raidr, liu2013experimental, keeth2001dram, mutlu2007stall, moscibroda2007memory, mutlu2008parbs, kim2010atlas, subramanian2014bliss, salp, kim2014flipping, qureshi2015avatar,
hassan2016chargecache, chang2016understanding, lee2017design,  chang2017understanding,  patel2017reaper,kim2018dram, kim2020revisiting, hassan2019crow, frigo2020trrespass, chang2014improving, chang2016low, vampire2018ghose, hassan2017softmc, khan2016parbor, khan2016case, khan2014efficacy, seshadri2015gather, seshadri2017ambit, kim2018solar, kim2019d, patel2019understanding, patel2020beer, lee2013tiered, lee2015decoupled, seshadri2013rowclone, luo2020clrdram, seshadri2019dram, wang2020figaro}}.

\subsection{DRAM Organization and Operation}
\begin{figure}[t]
    \centering
    \includegraphics[width=\linewidth]{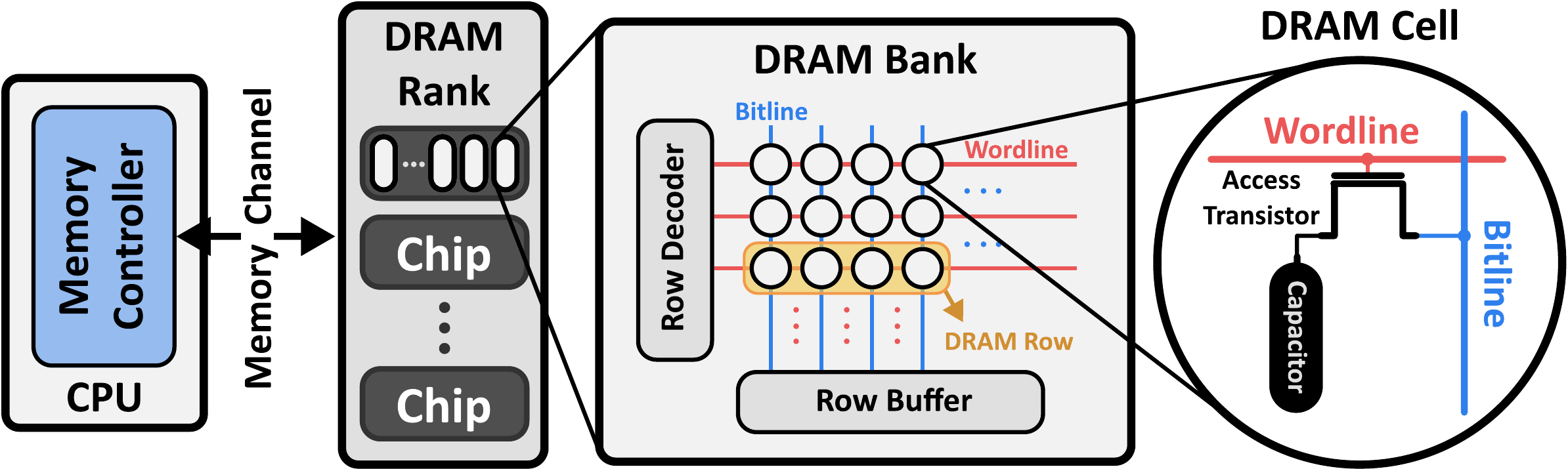}
    \caption{{Structure of a typical DRAM-based  system.}}
    \label{fig:dram_structure}
\end{figure}

Figure \ref{fig:dram_structure} shows the high-level structure of a typical DRAM{-based system}. At the lowest level of the hierarchy, DRAM {stores data within} \emph{cells} {that each consist} of a single capacitor and an access transistor. Each cell encodes a single bit of data using the ``high'' and ``low'' voltage states of the capacitor. Because a DRAM cell leaks charge over time, each cell's charge is periodically restored (i.e., refreshed) (e.g., every 32 or \SI{64}{\milli\second}~\cite{liu2012raidr, liu2013experimental, jedec2017, jedec2015low}) to prevent data loss.
{Cells are arranged in two-dimensional arrays to form DRAM \emph{banks}.}

{DRAM cells in a bank are addressed using \emph{rows} and \emph{columns}. A \emph{wordline} {drives} all DRAM cells in a \emph{row}, and a \emph{bitline} connects all DRAM cells {in} a \emph{column}.}
{All rows} within a bank share the peripheral circuitry, so only one row may be accessed per bank at any given time.
Each row begins in a \emph{closed} (i.e., \emph{precharged}) state and needs to be \emph{opened} (i.e., \emph{activated}) before any READ or WRITE operations can be performed {on it}.
Activating a row fetches the row's contents into the \emph{row buffer}.
The row buffer serves all read and write requests after fetching the data.
The row must be \emph{closed} before further accesses can be made to other rows of the same bank.

{A DRAM chip contains multiple banks that can be accessed in parallel. Multiple chips form a DRAM \emph{rank}.}
{At the highest level of the hierarchy, the \emph{memory controller} {in the {CPU die}} interfaces with {a DRAM rank} through a \emph{memory channel}. The memory controller serves memory access requests from various system components by issuing DRAM bus commands (e.g., activate, precharge, read, write, and refresh). The memory controller must schedule commands according to standardized \emph{timing parameters}, which are defined {in DRAM datasheets} to ensure that each operation has enough time to complete before starting the next~\cite{jedec2015low, jedec2017, jedec2015hbm, jedecddr, micron2014ddr4}.
The overall strategy that the memory controller uses to schedule commands is known as a \emph{scheduling policy.}}
{Typical {policies} seek} to optimize performance, fairness, quality of service (QoS), and energy {across applications running on a system}~\cite{rixner00, mutlu2008parbs, kim2010atlas, kim2010thread, ebrahimi2011parallel,  ausavarungnirun2012staged, subramanian2014bliss, subramanian2016bliss, usui2016dash}.
{Therefore, the} scheduling policy effectively controls all accesses to all DRAM channels, banks, rows, and columns.

\subsection{The RowHammer Phenomenon}

{RowHammer is a DRAM failure mode in which repeated activations to a single row (i.e., \emph{aggressor} row) cause disturbance capable of inducing bit-flips in {\emph{physically-nearby}} rows  (i.e., \emph{victim} rows) that are not being accessed~\cite{kim2014flipping}.}
{These bit-flips manifest after a row's activation count reaches a certain threshold {value} {within a refresh window}, which we call \emph{RowHammer threshold} (\nrh{}) {({{also} {denoted} 
as $MAC$}~\cite{jedec2017} and \smash{$HC_{first}$}~\cite{kim2020revisiting})}.}
Prior works {study the error characteristics of} RowHammer {bit-flips} and show that 
\nrh{} {varies across {DRAM} vendors, device models, generations, and chips}~\cite{kim2014flipping, park2016statistical,  frigo2020trrespass, kim2020revisiting, cojocar2020rowhammer}.
{Yang et al.~\cite{yang2019trap} explain this \nrh{} variation based on changing physical distances between adjacent wordlines (i.e., physical DRAM rows).}
Since DRAM chip density increases {at smaller feature sizes, both} Yang et al.'s observation and recent experimental studies~\cite{frigo2020trrespass, kim2020revisiting, kim2014flipping} {clearly demonstrate} that RowHammer {worsens with continued technology scaling~\cite{mutlu2017rowhammer, mutlu2019rowhammer}}. {In addition, recent studies {show} that emerging memory technologies also exhibit RowHammer vulnerability~\cite{khan2018sttramrowhammer, mutlu2017rowhammer, mutlu2019rowhammer}}.

\subsection{{{In-}DRAM Row Address Mapping}}
\label{sec:background_physicallayout}
DRAM vendors {often use {DRAM-internal} mapping} {schemes} to internally translate memory-controller-visible row addresses to physical row {addresses~\cite{kim2014flipping, smith1981laser, horiguchi1997redundancy, keeth2001dram, itoh2013vlsi, liu2013experimental,seshadri2015gather, khan2016parbor, khan2017detecting, lee2017design, tatar2018defeating, barenghi2018software, cojocar2020rowhammer,  patel2020beer} for two reasons:}
(1)~to optimize their chip design for density, performance, and power {constraints}; and (2)~{to improve factory yield by mapping the addresses of faulty rows to more reliable spare rows} {(i.e., post-manufacturing row repair).}
{Therefore, row {mapping} schemes can vary {with} (1) chip design {variation} across different vendors, DRAM models, and generations and (2) manufacturing process variation across different chips of the same design}. 
{State-of-the-art RowHammer mitigation mechanisms must account for {both sources of variation in order to}
be able to {accurately} identify all potential victim rows that are {physically nearby}
an aggressor row.}
{{Unfortunately}, DRAM vendors {consider their in-DRAM row address mapping schemes to be highly proprietary and ensure not to reveal mapping details in any public documentation because exposing the row {address} mapping {scheme} can reveal insights into the chip design and factory yield}~\cite{smith1981laser, horiguchi1997redundancy, keeth2001dram, itoh2013vlsi, lee2017design, patel2020beer}}.

%% file: 03_blockhammer.tex
\section{BlockHammer}
\label{sec:blockhammer}
BlockHammer is designed to (1) {scale efficiently as DRAM chips become increasingly vulnerable to RowHammer} and (2) be compatible with commodity DRAM chips. %
BlockHammer consists of two components.
The first component, RowBlocker {(\secref{rowblocker})}, prevents any possibility of a {RowHammer} bit-flip
by making it impossible to access a DRAM row at a high enough rate to induce RowHammer bit-flips. RowBlocker achieves this by efficiently tracking row activation rates using Bloom filters and throttling the row activations that target rows with high activation rates.
We implement RowBlocker entirely within the memory controller, ensuring RowHammer-safe operation without {any proprietary information about or modifications to the DRAM chip}. Therefore, RowBlocker is compatible with all commodity DRAM chips.
The second component, AttackThrottler (\secref{hammerthrottler}), alleviates the performance degradation a RowHammer attack can impose upon benign applications by selectively reducing the memory bandwidth usage of \emph{only} threads that AttackThrottler identifies {as likely} RowHammer attack{s} %
{(i.e., \emph{attacker threads}).
{By doing so, AttackThrottler provides a larger memory bandwidth to benign applications compared to a baseline system {that does not} throttle {attacker threads}. {As DRAM chips become more vulnerable to RowHammer,} AttackThrottler throttles {attacker threads} more aggressively, freeing even more memory bandwidth for benign applications to use.}
{By combining RowBlocker and AttackThrottler, BlockHammer achieves both of {its {design} goals}.}}

\subsection{RowBlocker}
\label{sec:mech_overview}
\label{sec:rowblocker}
RowBlocker's goal is to proactively throttle row activations {in an efficient manner} to avoid any possibility of a RowHammer attack. RowBlocker achieves this by overcoming two challenges regarding performance and area overheads.

\begin{figure}[ht!]
    \centering
    \includegraphics[width=0.48\textwidth]{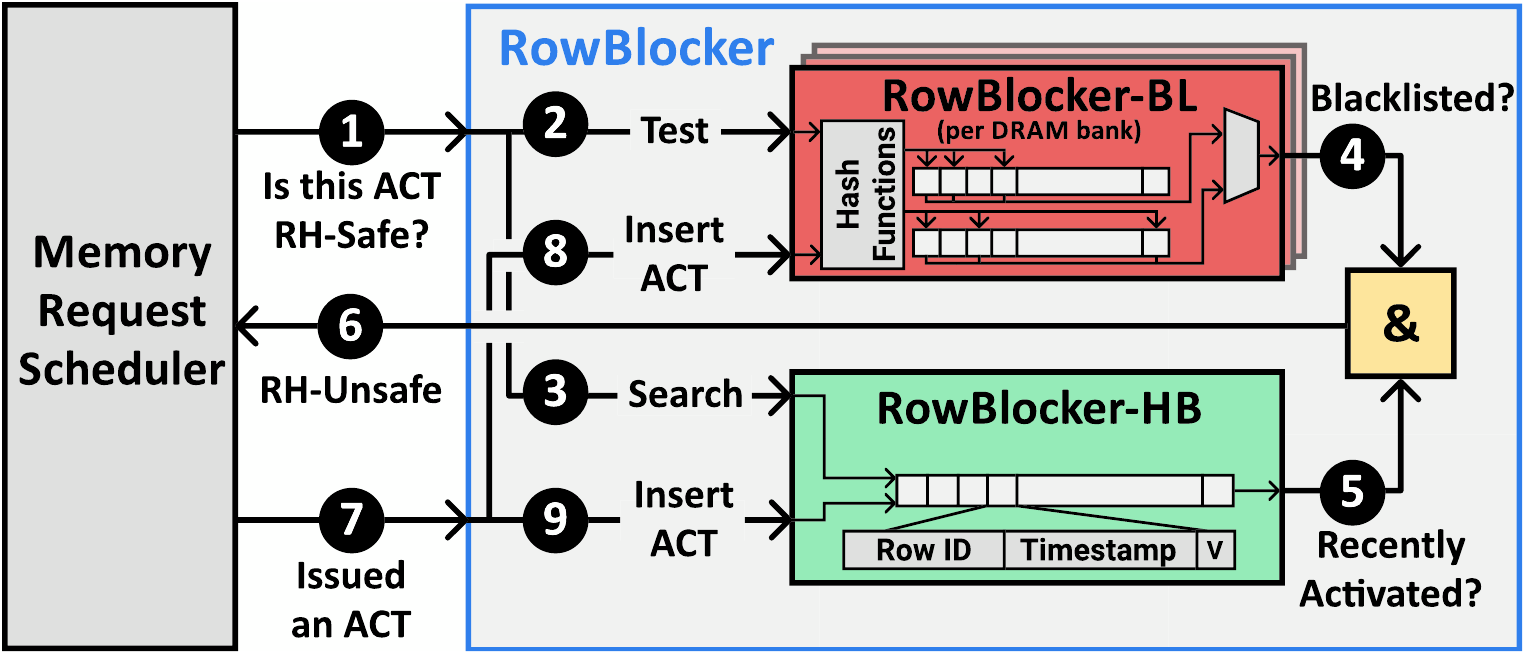}
    \caption{High-level overview of RowBlocker {(per DRAM rank)}. {An ACT is accompanied by its row address.}}
    \label{fig:overview}

\end{figure}

{First, achieving low performance {overhead} is a key challenge for a throttling mechanism because many benign applications tend to repeatedly activate a DRAM row that they have recently activated~\cite{kandemir2015memory, salp, hassan2016chargecache, hassan2019crow}. This can potentially {cause} a throttling mechanism to mistakenly throttle benign applications, thereby degrading system performance}.
{To ensure {throttling} \emph{only} applications that might cause RowHammer bit-flips,} RowBlocker throttles the row activations targeting \emph{only} rows whose activation rates are above a {given} threshold. To this end, RowBlocker implements two components as shown in \figref{overview}:
(1)~a {per-bank} blacklisting mechanism, \rbbl{}, which blacklists {all} rows with an activation rate {greater} than a predefined threshold called {the} \emph{blacklisting threshold} (\nbl{}); and
(2)~a {per-rank} activation history buffer, \rbhb{}, which tracks the most recently activated rows. RowBlocker enforces a time delay between two consecutive activations targeting a row \emph{only if} the row is \emph{blacklisted}.
By doing so, RowBlocker {is less likely to throttle} a benign application's row activations.

Second, achieving low area overhead is a key challenge for a throttling mechanism because throttling requires tracking all row activations {throughout} an entire refresh window \emph{without} losing information of any row activation. RowBlocker implements its blacklisting mechanism, \rbbl{}, by using area-efficient \emph{{counting Bloom filters}}~\cite{bloom1970space, fan2000summary} to track row activation rates. \rbbl{} maintains two counting Bloom filters in a {time-interleaved} manner to track row activation rates for large time windows without missing any row that should be blacklisted. We explain how counting Bloom filters work and how \rbbl{} employs them in \secref{mech_detect}.

\noindent
\textbf{High-Level Overview of RowBlocker.}
RowBlocker modifies the memory request scheduler to temporarily block {(i.e., delay)} an activation that targets a \emph{blacklisted} and {\emph{recently-activated}} row {until {the activation} {can be safely performed}}. {{By blocking such {row activations}}, RowBlocker ensures that} no row can be activated at a high enough rate to induce RowHammer bit-flips.
When the memory request scheduler attempts to schedule a row activation command {to a bank}, it queries RowBlocker (\circled{1}) to check if the row activation is RowHammer-safe. This simultaneously triggers two lookup operations. First, RowBlocker checks the \rbbl{} to see if the row to be activated is blacklisted (\circled{2}).
A row is blacklisted if {its activation rate exceeds}
a given threshold. We discuss how \rbbl{} estimates the activation rate of a row in \secref{mech_cbf}.
Second, RowBlocker checks \rbhb{} to see if the row has been recently activated (\circled{3}).
{If a row is both blacklisted (\circled{4}) \emph{and} recently activated (\circled{5}), RowBlocker responds to the memory request scheduler with {a} \emph{{RowHammer}-unsafe} signal (\circled{6}), consequently blocking the row activation.
Blocking such a row activation is essential because allowing further activations to a blacklisted and {recently-activated} row could increase the {row's} overall activation rate and thus result in RowHammer bit-flips.}
The memory request scheduler does \emph{not} issue a row activation
if RowBlocker returns \emph{unsafe}. {However, it} keeps issuing the \emph{RowHammer-safe} requests. {This scheduling {decision}}
effectively prioritizes {RowHammer-safe} memory accesses {over unsafe ones}.
{An unsafe row activation becomes safe again as soon as a certain amount of time (\tdelay{}) passes after its latest activation, {effectively limiting the row's average} {activation rate to a RowHammer-safe {value}.}
After \tdelay{} is satisfied, \rbhb{} no longer reports that the row {has been} recently activated (\circled{5}), thereby allowing the memory request scheduler to issue the row activation (\circled{6}).}
{When the memory request scheduler issues a row activation~(\circled{7}), it simultaneously updates both \rbbl{}~(\circled{8}) and \rbhb{}~(\circled{9}).
We explain how \rbbl{} and \rbhb{} work in \secref{mech_detect} and~\ref{sec:mech_prevent}, respectively.}

\subsubsection{{\rbbl{} Mechanism}}
\label{sec:mech_detect}
\label{sec:mech_cbf}
\label{sec:mech_switch}
\rbbl{} uses two counting Bloom filters (CBF) in a time-interleaved fashion to decide whether a row should be blacklisted. Each CBF takes turns to make the blacklisting decision.
A row is blacklisted when its activation rate exceeds a configurable threshold, which we call the \emph{blacklisting threshold} (\nbl{}).
When a CBF blacklists a row, {any further activations targeting the row} are throttled until the end of the CBF's turn.
In this subsection, we describe how a CBF works, how we use two CBFs to avoid stale blacklists, and how {the} two CBFs never fail to blacklist an aggressor row.

\noindent
\textbf{{Bloom Filter.}}
A Bloom filter~\cite{bloom1970space} is a space-efficient probabilistic data structure {that is} used for testing {whether} a set contains a particular element. %
A Bloom filter consists of a set of hash functions and a {bit array} {on which it performs} three operations: {\emph{clear}, \emph{insert}, and \emph{test}}.
Clearing a Bloom filter zeroes its {bit array}.
To insert/test an element,
each hash function evaluates an index
into the {bit array} for the element, {using an {identifier} for the element}.
Inserting {an element} sets the bits that the hash functions point to. Testing for {an element} checks whether all these bits are set.
Since a hash function can yield the same {set of} {indices} for different elements {(i.e., aliasing)}, testing a Bloom filter can return true for an element that was never inserted (i.e., false positive).
However, the \emph{test} {operation} never returns false for an inserted element (i.e., no false negatives).
A Bloom filter eventually saturates (i.e., {always returns true when tested for any element})
{if elements {are continually} inserted,}
{which requires periodically clearing the filter and losing all inserted elements.}

\noindent{}\textbf{Unified Bloom Filter (UBF).} {UBF~\cite{li2012compression} is} a Bloom filter variant that allows a system to continuously track a set of elements that are inserted into a Bloom filter within the most recent time window of a fixed length (i.e., a \emph{rolling time window}). {Using a conventional Bloom filter} to track a rolling time window could result in data loss whenever {the Bloom filter is} cleared, as the clearing eliminates the elements that still fall within the rolling time window.
{Instead,} {UBF} {continuously {tracks} insertions in a rolling time window by {maintaining}} \emph{two} Bloom filters {and using them in} a {time-interleaved} manner. {UBF {inserts every element into} both filters,}
while {the filters} take turns in responding {to} {\emph{test}} queries across consecutive limited time windows {(i.e., {\emph{epochs}})}. {UBF clears the filter which responds to {\emph{test}} queries at the end of an epoch and redirects the {\emph{test}} queries to the other filter for the next epoch. Therefore, each filter is cleared every {other epoch (i.e., the filter's lifetime is two epochs).}}
By doing so, UBF ensures no false negatives for the elements that are inserted in {a rolling time window of up to two epochs}.

\noindent
\textbf{{Counting Bloom Filter (CBF).}}
{To track}
\emph{the number of times} an element {is} inserted
into the filter, {another Bloom filter variant, called}
\emph{counting} Bloom filters {(CBF)}~\cite{fan2000summary},
{replaces} the {bit array} with a {\emph{counter} array}.
{Inserting an element in a CBF} \emph{increments} all of its corresponding counters. {T}esting an element returns the \emph{minimum} value {among} all {of {the element's}} {corresponding} counters, which represents an \emph{upper bound} on the number of times {an element} was inserted into the filter.
Due to aliasing, the test result can be \emph{larger} than the true insertion count, but it \emph{cannot} be {smaller} than that because counters are \emph{never decremented} (i.e., false positives are possible, but false negatives are not).

\noindent
\textbf{{Combining} {UBF and} CBF for Blacklisting.}
{To estimate row activation rates with low area cost, \rbbl{} combines {the ideas of} UBF and CBF to form our \emph{dual} counting Bloom filter (D-CBF). D-CBF maintains \emph{two} CBFs in the time-interleaved manner of UBF.}
{On every row activation, \rbbl{} inserts the {activated row's} address into {both} CBFs. \rbbl{} considers a row {to be} \emph{blacklisted} when {the row's} activation {count} exceeds {the blacklisting threshold} (\nbl{}) {in a rolling time window}.}

\begin{figure}[ht!]
    \centering
    \includegraphics[width=\columnwidth]{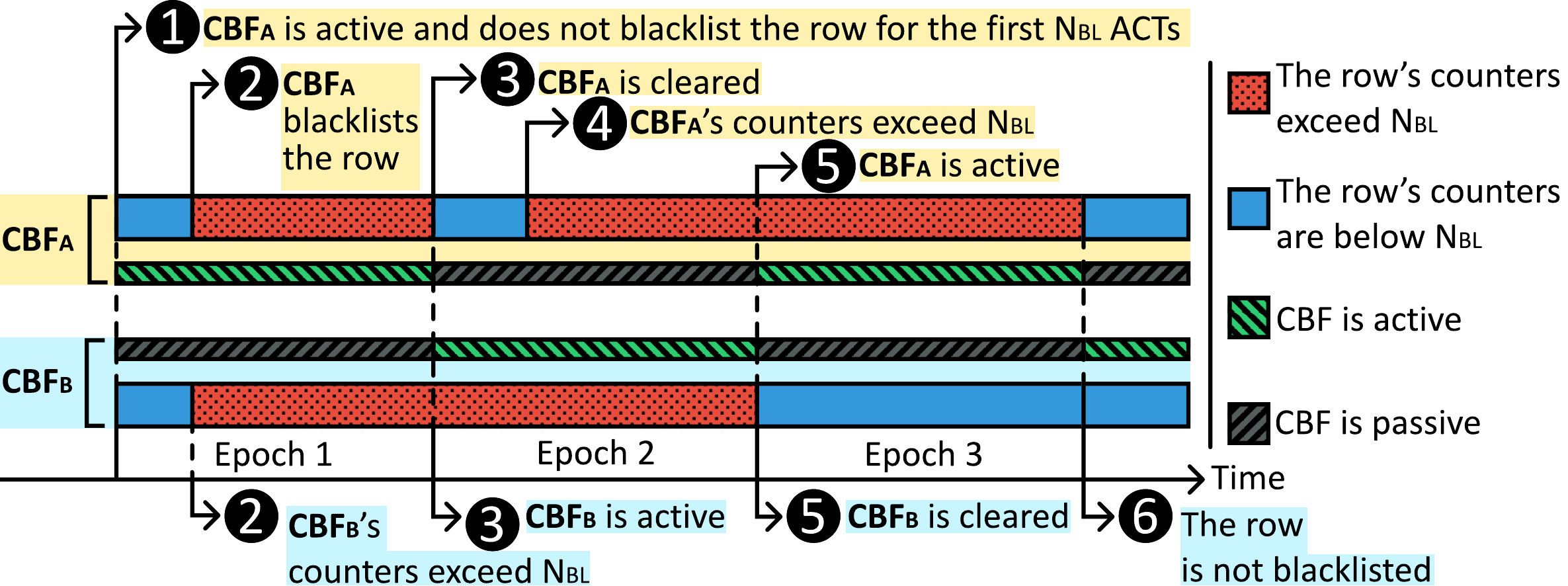}
    \caption{{{D-CBF} operation from a DRAM row's perspective}.}
    \label{fig:bloomfilter_timing}
\end{figure}

{\figref{bloomfilter_timing} illustrates how \rbbl{} uses a D-CBF over time.}
{\rbbl{} {designates one of the CBFs as \emph{active} and the other as \emph{passive}}.}
At {any given time,} only the \emph{active} CBF responds to {\emph{test}} queries.
When a {\emph{clear}} signal is received, D-CBF (1)~clears only the active filter (e.g., \cbfa{} at \circled{3}) and (2)~swaps the active and passive filters (e.g., \cbfa{} becomes passive and \cbfb{} becomes active at \circled{3}).
\rbbl{} blacklists a row if {the row's} activation count in the active CBF exceeds
the blacklisting threshold (\nbl{}).

\noindent
\textbf{{D-CBF Operation} Walk-Through.}
{{We walk through {D-CBF operation in} \figref{bloomfilter_timing}} from the perspective of a DRAM row.}
{The counters that correspond to {the row} in} both filters (\cbfa{} and \cbfb{}) are initially {zero} (\circled{1}).
\cbfa{} is the \emph{active} filter, while \cbfb{} is the \emph{passive} {filter}.
As the row's activation count accumulates and reaches \nbl{} (\circled{2}), both \cbfa{} and \cbfb{} decide to blacklist the row. RowBlocker applies the active filter's decision (\cbfa{}) and blacklists the row.
As the counter values do not decrease, the row remains blacklisted until the end of {Epoch~1}. Therefore, a minimum delay is enforced between consecutive activations {of} this row between \circled{2} and \circled{3}. At the end of {Epoch~1} (\circled{3}), \cbfa{} is cleared, and \cbfb{} becomes the active filter. Note that \cbfb{} immediately blacklists the {row,
as the counter values corresponding to the row in \cbfb{} are still larger than \nbl{}}. Meanwhile, assuming that the row continues to be activated, the counters in \cbfa{} {again} reach \nbl{} (\circled{4}). {At} the end of {Epoch~2} (\circled{5}), \cbfa{} becomes the active filter again and immediately blacklists the row. {By following this scheme, D-CBF blacklists the row as long as the row's activation count exceeds \nbl{} in an epoch.}
{Assuming that} the row's activation count does not exceed \nbl{} within Epoch~3, starting from \circled{6}, the row is {no longer blacklisted}.
Time-interleaving across {the} two CBFs ensures that BlockHammer maintains a \emph{fresh} blacklist {that} never incorrectly excludes a DRAM row that needs to be blacklisted.
\secref{mech_security} provides a generalized analytical proof of BlockHammer's security guarantees that comprehensively studies all possible row activation patterns across all epochs.

To prevent any specific row from being repeatedly blacklisted due to its CBF counters aliasing with those of an aggressor row (i.e., due to a false positive), \rbbl{} alters the hash functions that each CBF uses whenever the CBF is cleared.
{To achieve this, \rbbl{} replaces the hash function's seed value with a {new} random{ly-generated} value, as we explain {next}}.
Consequently, an aggressor row aliases with a different set of rows after every {\emph{clear}} operation.

\noindent
\textbf{Implementing Counting Bloom Filters.}
{To periodically send a \emph{clear} signal to D-CBF,} \rbbl{} implements a clock register that stores the timestamp of the {latest \emph{clear} operation}.
{In our implementation, each CBF} contains 1024 elements of 12-bit saturating counters to count up to the blacklisting threshold~\nbl{}. We employ four area- and latency-efficient H3-class hash functions that consist of {simple} static bit-shift and {mask}
operations~\cite{carter1979universal}. {We hardwire the static shift operation, so it does not require any logic gates. The mask} operation performs a {bitwise} exclusive-OR on the shifted element (i.e., row address) and a seed. To alter the hash function {when a CBF is cleared},
{RowBlocker simply replaces the hash function's seed value with a randomly-generated value}.

\subsubsection{{\rbhb{} Mechanism}}
\label{sec:mech_prevent}
\label{sec:mech_rahb}
{\rbhb{}'s goal is to ensure {that} a blacklisted row {cannot be} activated often enough to cause a bit-flip.}
{To ensure this,} {\rbhb{} delays a subsequent activation to a blacklisted row until the row's last activation becomes older than a certain amount of time that we call \tdelay{}.}
{To do so, \rbhb{} maintains a first-in-first-out history buffer that stores a record of all row activations in the last \tdelay{} time window.}
{When RowBlocker queries \rbhb{} with a row address (i.e., \circled{3} in Figure~\ref{fig:overview}), \rbhb{} searches the row address in the history buffer and {sets the} {\emph{``Recently Activated?''}} signal {to true} if the row address appears in the history buffer.}

\noindent
\textbf{Implementing \rbhb{}.}
{{We implement} a per-DRAM-{rank} history buffer as a circular queue using a head and a tail pointer. Each entry of this buffer stores (1)~a row {{ID} (which is unique in the rank)}, (2)~a timestamp of when the entry was inserted into the buffer, and (3)~a valid bit.}
{The head and the tail pointers address the oldest and the youngest entries in the history buffer, {respectively}.} {When the memory request scheduler issues a row activation (\circled{7} in \figref{overview}), \rbhb{} inserts a new entry with the activated row address, the current timestamp, and a valid bit {set to} {logic `1'} into the history buffer and updates the tail pointer.}
{\rbhb{} checks the timestamp of the oldest entry, {indicated by the head pointer}, every cycle. When the oldest entry becomes as old as \tdelay{}, \rbhb{} invalidates {the entry} by resetting its valid bit to {logic `0'} and updates the head pointer.}
{To test whether a row is recently activated (\circled{3} in \figref{overview}), \rbhb{} looks up the tested row address in each \emph{valid} entry (i.e., {an} entry with a valid bit set to one) in parallel. {To search the history buffer {with} low latency, we keep row addresses in a content-addressable memory array.}
Any matching \emph{valid} entry means that} the row has been activated {within the last \smash{\tdelay{}} time window, so}
the {new} activation should not be issued if the row is blacklisted by \rbbl{}.
We {size} the {history buffer} {to be} large enough to contain the {worst-case} number of row
activations that need to be {tested}.
{The number of activations that can be performed in a DRAM rank is bounded by {the} timing parameter \tfaw{}~\cite{jedec2017, jedec2015hbm, jedec2015low, micron2014ddr4}, which defines a rolling time window that can contain {at most} four row activations. Therefore, within a \smash{\tdelay{}} time window{,} there can be at most \smash{$\lceil4\times\tdelayn{}/\tfawn{}\rceil$} row} activations.

\noindent
\textbf{{Determining How Long to Delay an Unsafe Activation.}}
To avoid RowHammer {bit-flips}, a row's activation {count should not exceed the RowHammer threshold (\nrh{})}
{within a refresh window (\trefw{}).}
RowBlocker satisfies this upper bound activation rate within each CBF's {lifetime (\tbf{}), which is the time window between two \emph{clear} operations applied to a CBF (e.g., Epochs 1 and 2 for \cbfb{} and Epochs 2 and 3 for \cbfa{} in \figref{bloomfilter_timing}).}
{To ensure an upper bound activation rate of \nrh{}/\trefw{} at all times, RowBlocker does not allow a row to be activated more than $(\tbfn{}/\trefwn{})\times\nrhn{}$ times within a \tbf{} time window.}
In the worst-case {access pattern within a CBF's lifetime,} a row is activated \nbl{} times at the very beginning of the \tbf{}
time window {as rapidly as possible}, taking a total time of $\nbln{} \times \trcn{}$. In this case, RowBlocker evenly distributes the activations that it can allow (i.e., $(\tbfn{}/\trefwn{})\times\nrhn{}-\nbln{}$) {throughout} the rest of the window (i.e., $\tbfn{} - (\nbln{} \times \trcn{})$).
Thus, we define \tdelay{} as shown in Equation~\ref{equ:tdelay}.
\begin{equation}
  \tdelayn{} = \frac{\tbfn{} - (\nbln{} \times \trcn{})}{(\tbfn{}/\trefwn{})\times\nrhn{} - \nbln{}}
  \label{equ:tdelay}
\end{equation}
\subsubsection{Configuration}
{RowBlocker has} three {tunable} configuration parameters {that collectively define RowBlocker's false positive {rate} and area characteristics}: (1)~{the CBF} size{:} the number of counters in a CBF; (2)~\tbf{}: the CBF lifetime; and (3)~\nbl{}: the blacklisting threshold.
{Configuring the CBF size directly impacts the CBF's area and false positive rate (i.e., the fraction of mistakenly blacklisted row activations) because the CBF size determines {both} the CBF's physical storage requirements and the likelihood of unique row addresses aliasing to the same counters. Configuring \nbl{} and \tbf{} determines the penalty of each false positive and the area cost of \rbhb{}'s history buffer, because \nbl{} and \tbf{} jointly determine the delay between activations required for RowHammer-safe operation (via Equation~\ref{equ:tdelay}) and the maximum number of rows that RowBlocker must track within each epoch.}

To determine suitable values for each of the three parameters, we follow a three-step methodology that minimizes the cost of false positives for a given area budget. First, we empirically choose the CBF size based on false positive rates observed in our experiments (\secref{methodology} discusses our experimental configuration).
{We choose a CBF size of 1K counters {because} we observe that reducing the CBF size below 1K {significantly} increases the false positive rate due to aliasing.}

{Second, we {configure} \nbl{} {based on} three goals: (1)~\nbl{} should be smaller than the RowHammer threshold to prevent RowHammer bit-flips; (2)~\nbl{} should be significantly {larger} than the per-row activation counts that benign applications exhibit {in order to ensure that RowBlocker does not blacklist benign applications' row activations, even when accounting for false positives due to Bloom filter aliasing;}
and (3)~\nbl{} should be as low as possible to {minimize} \smash{\tdelay{}} ({i.e., the time delay penalty for all activations to blacklisted rows, including those due to false positives) per \equref{tdelay}.}}
To {balance these three goals,} we analyze the memory access patterns of 125 eight-core multiprogrammed {workloads}, each of which consists of eight {randomly-chosen} benign {threads}.
We simulate these {workloads} {using {cycle-level} simulation~\cite{Kim2016Ramulator, ramulatorgithub}} for 200M instructions with a warmup {period} of 100M instructions on a \SI{3.2}{\giga\hertz} system with \SI{16}{\mega\byte} of last-level cache. We measure per-row activation rates by counting the activations that each row experiences within {a} \SI{64}{\milli\second} time window {(i.e., one refresh window)} starting {from} the row's first activation. We observe that benign {threads} reach up to 78, 109, and 314 activations per row in a \SI{64}{\milli\second} time window for {the} 95th, 99th, and 100th percentile of the set of DRAM rows that are accessed at least once.
{Based on these observations}, we set \nbl{} to 8K for a RowHammer threshold of 32K, providing {(1)~RowHammer-safe operation, (2)~an} ample margin for row activations from benign threads {to achieve a low false positive rate (less than 0.01\%, as shown in \secref{evaluation_tech_scaling}), and (3)~a} reasonable {worst-case} \tdelay{} penalty {of {\SI{7.7}{\micro\second}}} for activations to blacklisted rows.

Third, we use Equation~\ref{equ:tdelay} to choose a value for \tbf{} such that the resulting {\smash{\tdelay{}}} does not excessively penalize a mistakenly blacklisted row (i.e., a false positive).
{Increasing \tbf{} both (1)~decreases \smash{\tdelay{}} (via \equref{tdelay}) and (2)~extends the length of time for which a row is blacklisted. Therefore, we set \tbf{} equal to \trefw{}, which achieves as low a {\tdelay{}} as possible without blacklisting a row past the point at which its potential victim rows have already been refreshed.}

{We present} the final values we choose for all BlockHammer parameters {in conjunction with the DRAM timing parameters we use}
in Table~\ref{table:tuned_params} after explaining how BlockHammer addresses {many-sided RowHammer attacks in \secref{many_sided_attacks}}.

\noindent
\textbf{Tuning for Different DRAM Standards.} The values in Table~\ref{table:tuned_params} depend on {three} timing constraints defined by the memory standard: (1)~the minimum delay between activations to the same bank (\trc{}), (2)~the refresh window (\trefw{}), and (3)~the four-activation window (\tfaw{}). The delay enforced by BlockHammer (\smash{\tdelay{}}) scales linearly with \trefw{}, while it is marginally affected by \trc{} (\equref{tdelay}). \trefw{} remains constant at \SI{64}{\milli\second} across DDRx standards from  DDR~\cite{jedecddr} to DDR4~\cite{jedec2017}, while \trc{} has marginally reduced from \SI{55}{\nano\second} to \SI{46.25}{\nano\second}~\cite{micron2014networking, jedec2015low, jedec2017, micron2014ddr4, jedec2015hbm}. Therefore, \smash{\tdelay{}} increases only marginally across several DDR generations. In LPDDR4, \trefw{} is halved, which allows a reduction in {\smash{\tdelay{}}, and thus the} latency penalty of a blacklisted row. \tfaw{} affects {only} the size of the history buffer, {and its value varies} between \SIrange{30}{45}{\nano\second} across modern DRAM standards~\cite{micron2014networking, jedec2015low, jedec2017, micron2014ddr4, jedec2015hbm}.

\subsection{AttackThrottler}
\label{sec:hammerthrottler}

{AttackThrottler's goal is to mitigate} the system-wide performance degradation that a RowHammer attack could {inflict upon benign applications.
AttackThrottler achieves this by using memory access patterns to {(1)}~identify and {(2)}~throttle threads that potentially {induce} a RowHammer attack.}
{First, to identify potential {RowHammer attack} threads, AttackThrottler exploits the fact that a {RowHammer attack} thread inherently attempts to issue more activations to a blacklisted row than a benign application would.}
{Thus}, AttackThrottler tracks {the exact number of} times each thread {performs a row activation to a {blacklisted row in each bank}.}
{Second, AttackThrottler applies a quota to the total number of in-flight memory requests allowed for {\emph{any}} thread that is {identified} to be a {potential} attacker (i.e., that frequently activates blacklisted rows). {Because} {such} a thread activates blacklisted rows more often, AttackThrottler reduces the thread's quota, reducing its memory bandwidth utilization. Doing so frees up}
{memory resources for concurrently-running benign applications that are \emph{not} repeatedly activating (i.e., hammering) blacklisted rows.}

\subsubsection{{Identifying Ongoing RowHammer Attacks}}

AttackThrottler {identifies {threads} that {exhibit} memory access patterns {similar to a RowHammer attack} by monitoring a new metric called the \emph{RowHammer likelihood index} (\rhli{}), which {quantifies} {the similarity between a given thread's memory access pattern and a real RowHammer attack}.}
{AttackThrottler calculates \rhli{} for each {<}thread, DRAM bank{>} pair. {\rhli{} is defined} as the number of {blacklisted row activations} the thread performs {to} the DRAM bank, normalized to the maximum
number of times a blacklisted row can be activated in a BlockHammer-protected system. As we describe in \secref{rowblocker}, a row's activation count {during one CBF} lifetime is bounded by the RowHammer threshold, scaled to a CBF's lifetime (i.e., $\nrhn{} \times (\tbfn{}/\trefwn{})$). Therefore, a blacklisted row that {has already been} activated \nbl{} times cannot be activated more than {$\nrhn{}\times(\tbfn{}/\trefwn{})-\nbln{}$} times. Thus, {AttackThrottler} calculates \rhli{} {as shown in} Equation~\ref{equ:rhli}, {during a CBF's lifetime}.}
\begin{equation}
  RHLI = \frac{Blacklisted~Row~Activation~Count}{\nrhn{}\times(\tbfn{}/\trefwn{})-\nbln{}}
  \label{equ:rhli}
\end{equation}
{The} {\rhli{} {of a {<}thread, bank{>} pair is 0 when a thread certainly does \emph{not} perform {a} RowHammer attack on the bank. As a {<}thread, bank{>} pair's \rhli{} {reaches} 1, the thread {is} more likely to induce RowHammer bit-flips in the bank.

\rhli{} never exceeds 1 in a BlockHammer-protected system because
AttackThrottler completely blocks a thread's memory accesses {to} a bank (i.e., applies a quota of zero to them) when the {<thread, bank>} pair's \rhli{} reaches 1, {as we describe in \subsubsecref{ht_throttling}}.}}
{\rhli{} can be used independently from BlockHammer as a metric quantifying a thread's potential to be a RowHammer attack, as we discuss in \secref{rhli_system_level}}.

To demonstrate example \rhli{} values, we conduct {cycle-level} simulations on a set of 125~{multiprogrammed workloads}, each of which consists of one RowHammer attack thread and seven benign threads {randomly-selected from the set of workloads we describe in \secref{methodology}.}
We measure {the \rhli{} values} of benign {threads and RowHammer attacks} {for BlockHammer's two modes: (1)~\emph{observe-only} and (2)~\emph{full-functional}.
In \emph{observe-only} mode, BlockHammer computes \rhli{} but does not interfere with memory requests. In this mode, only RowBlocker's blacklisting logic (\rbbl{}) and AttackThrottler's counters are functional, allowing BlockHammer to blacklist row addresses and measure \rhli{} per thread {without blocking any row activations}. In \emph{full-functional} mode, BlockHammer operates normally, i.e., {it} detects {the threads performing RowHammer attacks}, throttles their requests, and ensures {that} no row's activation rate exceeds {the} RowHammer threshold.}
{We set the blacklisting threshold to {512}~activations} in a \SI{16}{\milli\second} time window.
{We make two observations from these experiments. First, benign applications exhibit zero \rhli{} because their row activation counts {never exceed} the blacklisting threshold. {On the other hand,} RowHammer attacks reach an average (maximum, minimum) \rhli{} value of 10.9 (15.5, 6.9) in observe-only mode, showing that {an \rhli{} greater than 1 reliably {distinguishes} a} RowHammer attack thread. Second, when in full-functional mode, BlockHammer {reduces an attack's \rhli{} by 54x on average, effectively reducing the \rhli{} of all RowHammer attacks to below 1.}} {BlockHammer does not affect} {benign applications' \rhli{} values, which stay at zero}.

AttackThrottler calculates \rhli{} separately for each {<thread, bank>} pair. {To do so,}
{AttackThrottler maintains two counters per {<thread, bank>} pair, using the same time-interleaving mechanism as the dual counting Bloom filters (D-CBFs) in RowBlocker (see \secref{mech_detect}).
At any given time, one of the counters is designated as the active counter, while the other is designated as the passive counter. Both counters are incremented when {the thread activates a} blacklisted row {in the bank}. {Only the} active counter is used to calculate \rhli{} {at any point in time}.
{When} RowBlocker clears its active filter {for a given bank}, AttackThrottler clears {each thread's active counter {corresponding to} the bank and swaps the active and passive counters.}

We implement AttackThrottler's counters as saturating counters}
because \rhli{} never exceeds 1 in a BlockHammer-{protected} system. Therefore, a{n} AttackThrottler counter saturates at the RowHammer threshold normalized to a CBF's lifetime, which we calculate as $\nrhn{}\times(\tbfn{}/\trefwn{})$.
{For the configuration we provide in Table~\ref{table:tuned_params}}, AttackThrottler's counters require {only} four bytes of additional storage in the memory controller for each {<thread, bank>} pair
{(e.g., 512~bytes in total for an eight-thread system with a 16-bank DRAM rank).}

\subsubsection{{Throttling {RowHammer} Attack {Threads}}}
\label{subsubsec:ht_throttling}
AttackThrottler throttles any thread  with {a non-zero} \rhli{}.
{To do so, AttackThrottler limits the in-flight request count of each {<thread, bank>} pair by applying a quota inversely proportional to the {<thread, bank>} pair's \rhli{}.}
{Whenever a thread reaches its quota, the thread is {\emph{not}} allowed to make} {a} new memory request {to the shared caches} {or directly to the main memory} until {one of} its in-flight requests {is} completed.
If the thread continues to activate blacklisted rows {in a bank, its \rhli{} increases and consequently} its quota decreases{.}
This slows down the {RowHammer attack thread} while freeing up additional memory bandwidth for {concurrently-running} benign threads that experienc{e} no throttling due to their {zero} \rhli{}. In this way, BlockHammer mitigates the performance overhead that a RowHammer attack could inflict upon benign applications.

\subsubsection{{Exposing RHLI to the System Software}}
\label{sec:rhli_system_level}
{Although BlockHammer operates independently from the system software, e.g., the operating system (OS), BlockHammer can optionally expose its {per-DRAM-bank, per-thread \rhli{} values} to the OS. The OS can then use {this information} to mitigate
an ongoing
RowHammer attack at the software level. For example, the OS might kill or deschedule an attacking thread to prevent it from negatively impacting the system's performance and energy.} We leave {the study of OS-level mechanisms} using \rhli{} for future work.

%% file: 04_many_sided_attacks.tex
\section{Many-Sided RowHammer Attacks}
\label{sec:many_sided_attacks}
Hammering an aggressor row can disturb physically nearby rows even if they are not {immediately} adjacent~\cite{kim2014flipping, kim2020revisiting}, allowing {\emph{many-sided}} attacks that hammer {\emph{multiple}} DRAM rows to induce RowHammer bit-flips as a result of their cumulative disturbance~\cite{frigo2020trrespass}. Kim et al.~\cite{kim2014flipping} report that an aggressor row's impact decreases based on its physical distance to the victim row (e.g., by an order of magnitude per row) and disappears after a certain distance (e.g., 6 rows~\cite{kim2014flipping, frigo2020trrespass, kim2020revisiting}).

To address many-sided RowHammer attacks, we conservatively add up the effect of each row to reduce BlockHammer's RowHammer threshold (\nrh{}), such that the cumulative effect of concurrently hammering each row \nrhtuned{} times becomes equivalent to hammering only an immediately-adjacent row \nrh{} times.
{We} calculate \nrhtuned{} {using} three parameters:
(1)~\nrh{}: the RowHammer threshold for hammering a single row; (2)~{blast radius ($r_{blast}$)}: the
{maximum physical distance (in terms of rows) from the aggressor row at which RowHammer bit-flips can be observed;}
and (3)~{blast impact factor {($c_{k}$)}}: {the} ratio {between the} activation counts required to {induce a bit-flip in a victim row {by hammering {($i$)~an} immediately-adjacent row and {($ii$)~a} row} at {a} distance {of} $k$ rows away{.}} {W}e calculate the disturbance {that} hammering a row $N$ times {causes for} a victim row {that is physically located} $k$ {rows away as:} $N \times c_{k}$.
{\equref{tai_naggr} shows how we calculate \nrhtuned{} in terms of \nrh{}, {$c_{k}$}, and $r_{blast}$.
{We set \nrhtuned{} such that, even when all rows within the blast radius of a victim row (i.e., $r_{blast}$ rows on both sides of the victim row) are hammered for \nrhtuned{} times, their cumulative disturbance (i.e., $2\times(\nrhtunedn{}\times c_{1}+\nrhtunedn{}\times c_{2}+...+\nrhtunedn{}\times c_{r_{blast}})$) on the victim row will not exceed the disturbance of hammering an immediately-adjacent row \smash{\nrh{}} times.}}
\begin{equation}
\footnotesize
\nrhtunedn{} = \frac{\nrhn{}}{2\sum_{1}^{r_{blast}}{c_{k}}}, \quad \mathrm{where}
\begin{cases}
      c_{k} = 1, & \text{if $k = 1$}\\
      0 < c_{k} < 1, & \text{if $r_{blast} \geq k > 1$ }\\
      c_{k} = 0, & \text{if $k > r_{blast}$}
\end{cases}
\label{equ:tai_naggr}
\vspace{-0.5\baselineskip}
\end{equation}
{$r_{blast}=6$ and {\smash{$c_{k}={0.5}^{k-1}$}} are the {worst-case} values observed {in modern DRAM chips} based on experimental results presented in prior characterization studies~\cite{kim2014flipping, kim2020revisiting}, {which} characterize {more than 1500} real DRAM chips from different vendors, standards, and generations from {2010} to 2020. To support {a} DRAM chip {with {these} worst-case characteristics},}
we find that $\nrhtunedn{}$ should equal $0.2539\times\nrhn{}$ using \equref{tai_naggr}.
Similarly, to {configure BlockHammer for} double-sided attacks {{(which is} the attack model that state-of-the-art RowHammer mitigation mechanisms address}~\cite{kim2014flipping, son2017making, you2019mrloc, seyedzadeh2018cbt, lee2019twice, park2020graphene}{)}, we calculate \nrhtuned{} as half of \nrh{}
(i.e., {$r_{blast} = c_{k} = 1$}).
{Table~\ref{table:tuned_params}
presents BlockHammer's configuration
for timing specifications of a commodity DDR4 DRAM chip~\cite{jedec2017} and
a realistic RowHammer threshold of 32K~\cite{kim2020revisiting}, tuned to address double-sided attacks.}

\begin{savenotes}
\begin{table}[ht]
\footnotesize
\renewcommand{\arraystretch}{0.9}
\setlength{\tabcolsep}{3pt}
\centering
\begin{tabular}{l|llllll}
\textbf{Component} & \multicolumn{5}{l}{\textbf{Parameters}} \\ \hline
\multirow{2}{*}{DRAM Features} & \nrh{} &: {32K} & Banks &: 16 & \trc{}  &: \SI{46.25}{\nano\second}  \\
                               & {\nrhtuned{}} &: {16K} & \trefw{} &: \SI{64}{\milli\second}  & \tfaw{}  &: \SI{35}{\nano\second} \\
\hline
\multirow{3}*{\setlength\tabcolsep{0pt}\begin{tabular}{l}{\rbbl{}}\end{tabular}} & \nbl{}  &: 8K  & \tbf{} &: {\SI{64}{\milli\second}\quad} & \tdelay{}\footnote{This is the {theoretical} maximum delay that a row activation can {experience}.
Benign workloads {actually} experience {smaller delays of} up to \SI{1.7}{\micro\second}, \SI{3.9}{\micro\second}, {and \SI{7.6}{\micro\second} for P50, P90, and P100} of the row activations (see Section~\ref{sec:evaluation_false_positives}).} &: {\SI{7.7}{\micro\second}} \\
 & \multicolumn{2}{l}{CBF size}   & \multicolumn{4}{l}{: {1K} counters per CBF} {(per-bank)} \\
& \multicolumn{2}{l}{CBF Hashing} & \multicolumn{4}{l}{: 4 H3-class functions~\cite{carter1979universal} per CBF} \\
\hline
{\rbhb{}} & \multicolumn{2}{l}{Hist. buffer size} & \multicolumn{4}{l}{: {887} entries per {rank (16 banks)}} \\
\hline
AttackThrottler & \multicolumn{6}{l}{2 counters per <thread, bank> pair}\\
\hline
 \end{tabular}
 \caption{{Example BlockHammer parameter values based on DDR4 specifications~\cite{jedec2017} and RowHammer vulnerability~\cite{kim2020revisiting}.}}
 \label{table:tuned_params}
\end{table}
\end{savenotes}

%% file: 05_security_analysis.tex
\section{Security Analysis}
\label{sec:mech_security}
{W}e use the \emph{proof by contradiction} method to prove that no RowHammer attack can defeat BlockHammer {(i.e.,} activate a DRAM row more than \nrh{} times in a refresh window{)}. To do so, we begin with the assumption that there exists an access pattern that can exceed \nrh{} by defeating BlockHammer. Then, we mathematically represent all possible distributions of row activations and define the constraints for activating a {row more} than \nrh{} times in a refresh window. Finally, we show that it is impossible to satisfy these constraints, and thus, no {such} access pattern that can defeat BlockHammer {exists}. Due to space constraints, we briefly summarize all steps of the proof. We {provide the complete proof in an extended version}~\cite{yaglikci2020blockhammerarxiv}.

\noindent
\textbf{Threat Model.} We assume a {comprehensive} threat model in which the attacker can (1)~fully utilize memory bandwidth, (2)~precisely time each memory request, and (3)~comprehensively {and accurately} know details of the memory controller, BlockHammer, and DRAM implementation.
{In addressing this threat model,
{we do}
not consider any hardware or software component {to be} \emph{trusted} or \emph{safe} except {for} the memory controller, {the} DRAM chip, and the physical interface between those two.}

\noindent
\textbf{Crafting an Attack.}
We model a generalized {memory} access pattern that a RowHammer attack can exhibit from the perspective of an aggressor row. We represent {an attack's}
{row activation pattern in}
a series of epochs, each of which is bounded by RowBlocker's D-CBF {\emph{clear}} commands {to either CBF (i.e., half of the CBF lifetime or $\tbfn{}/2$), }%
as shown in \figref{bloomfilter_timing}.
According to the {time-interleaving mechanism} {(explained in Section~\ref{sec:mech_detect})}, the active CBF blacklists a row based on the row's total activation count in the current and previous epochs {to} limit the number of activations to the row. To demonstrate {that} RowBlocker effectively limits the number of activations to a row, {and therefore prevents} {all possible RowHammer attacks,}
we model all possible activation patterns targeting a DRAM row at the granularity of a single epoch. From the perspective of a CBF, each epoch can be classified based on the number of activations that the aggressor can receive in the previous (\npre{}) and current (\nagg{}) epochs. We identify five possible epoch types (i.e., $T_{0}-T_{4}$), which we list in Table~\ref{table:security_accpatterns}. {The table} {shows} (1)~the {range of row} activation counts in the previous epoch (\npre{}), (2)~the {range of row} activation counts in the current epoch (\nagg{}), {and (3)~the maximum {possible} row activation count in the current epoch ($\naggn{}_{max}$).}

\begin{table}[h]
 \footnotesize
\centering
\resizebox{\linewidth}{!}{
 \begin{tabular}{c|ccl}
  Epoch Type    & \npre{}        &           \nagg{}                  & $\naggn{}_{max}$ \\ %
  \hhline{=|===}
  $T_{0}$ &                &        $\naggn{} < \nblpn{}$       & $\nblpn{} - 1$      \\ %
  $T_{1}$ & $  < \nthn{} $ & $\nblpn{} \leq \naggn{} < \nthn{}$ & $\nthn{} - 1$    \\ %
  $T_{2}$ &                &      $\naggn{} \geq \nthn{}$       & $\tepochn{}/\tdelayn{} - (1-\trcn{}/\tdelayn{})\nblpn{}$    \\  \hline
  $T_{3}$ & \multirow{2}{*}{$\geq \nbln{}$} &      $ \naggn{} < \nbln{} $        & $\nbln{} - 1$    \\ %
  $T_{4}$ &                &      $\naggn{} \geq \nthn{}$       & $\tepochn{}/\tdelayn{} $    \\ \hline
 \end{tabular}
 }
 \caption{{Five possible epoch types that span all possible memory access patterns, {defined by the number of row activations the aggressor row can receive in the previous epoch (\npre{}) and in the current epoch (\nagg{}). $\naggn{}_{max}$ shows the maximum value of \nagg{}.}}}
 \label{table:security_accpatterns}
\end{table}

{{The epoch type indicates the recent activation rate of the aggressor row, and RowBlocker uses this information to determine whether {or not} to blacklist} the aggressor row {in the current and next epochs}. A $T_0$ epoch indicates that the row was activated {fewer} than $\nbln{}$ times in the previous epoch (i.e., $\npren{} < \nbln{}$) and {fewer} than $\nbln{}-\npren{}$ times (denoted as \nblp{} for simplicity) in the current epoch. Since the row was activated fewer times than the {blacklisting} threshold, the row {is} not blacklisted in the current epoch. Compared to $T_0$, a $T_1$ epoch indicates that the row was activated greater than \nblp{} times but fewer than \nbl{} times in the current epoch. Since the activation count exceeds the threshold \nblp{} but not \nbl{}, the row is blacklisted in the current epoch. {When a $T_1$ type epoch finishes, the row starts the next epoch as \emph{not blacklisted} because the row's activation count is {lower} than \nbl{}.} Compared to $T_1$, a $T_2$ epoch indicates that the row's activation count in the current epoch exceeds \nbl{}. Since the activation count exceeds the {blacklisting threshold} \nbl{}, the row is blacklisted in \emph{both} the current and next epochs.}

{A $T_3$ epoch indicates that the row's activation count in the previous epoch exceeded \nbl{} and the row is activated fewer times than \nbl{} times in the current epoch. In this case, the row is blacklisted in the current epoch, but no longer blacklisted in the {beginning of the} next epoch. Compared to $T_3$, a $T_4$ epoch indicates that the row is activated more than \nbl{} times in the current epoch. The row is blacklisted in both current and next epochs, as its activation rate is too high and {could} lead to a successful RowHammer attack {if not blacklisted}.}

{We calculate the upper bound for the total activation count an attacker can reach during the current epoch (shown under \nepmax{} in  Table~\ref{table:security_accpatterns}).} In {the} $T_{0}$, $T_1$, or $T_3$ {epochs}, by definition, a row's activation count cannot exceed $\nblpn{}-1$, $\nbln{}-1$, and $\nbln{}-1$, respectively.
{{In {a $T_{4}$ epoch}, the row is} already blacklisted from the beginning ($N_{0} \geq \nbln{}$). Therefore, {the row can be activated at most once in every \tdelay{} time window, resulting in an upper bound activation count of $\tepochn{}/\tdelayn{}$.}
{In {a $T_{2}$ epoch}, a row can be activated} \nblp{} times at a time interval as small as \trc{}, which takes $t_{1}=\nblpn{}\times\trcn{}$ time. Then, the row is blacklisted and further activations are performed with a minimum interval of \tdelay{}, which takes $t_{2}=(\nepmaxin{}-\nblpn{})\times\tdelayn{}$ time. Since all {of} these activations need to fit into the epoch's time window, we solve the equation $\tepochn{}=t_{1}+t_{2}$ for \nep{}, and derive \nepmax{} for an epoch of type $T_{2}$ as shown in Table~\ref{table:security_accpatterns}.}

\noindent
\textbf{Constraints of a Successful RowHammer Attack.}
{\lineskip=0pt\lineskiplimit=-\maxdimen{%
We mathematically represent a hypothetically successful RowHammer attack as a permutation of many epochs. We denote the number of instances for an epoch type $i$ as $n_i$ and the maximum activation count the epoch $i$ can reach as $\nepmaxin(i)$.
{To be successful, the RowHammer attack must satisfy} three constraints, {which} we present in Table~\ref{table:security_constraints}.}
(1)~{The attacker should activate a}n aggressor row more than \nrh{} times within a refresh window (\trefw{}).
(2)~Each epoch type can be preceded only by a {subset} of epoch types.\footnote{{Since we define epoch types based on activation counts in both the previous and current epochs, we note that consecutive epochs are dependent and therefore limited: an epoch of type $T_{0}$, $T_1$, or $T_2$ can be preceded only by an epoch of type $T_{0}$, $T_1$, or $T_3$, while an epoch of type $T_3$ or $T_4$ can be preceded only by an epoch of type $T_2$ or $T_4$}.}
Therefore,
an epoch type $T_x$ {cannot occur more} times {than} the total number of {instances of all epoch types that can {precede} epoch type $T_x$.}
(3)~An epoch cannot occur for a negative number of times.
}%

\begin{table}[h]
 \footnotesize
 \centering

 \begin{tabular}{rll}
 \hline
{(1)} & $\nrhn{} \leq \sum{(n_{i}\times\nepmaxin{})}$, & $\trefwn{} \geq \tepochn{}\times\sum{n_{i}}$\\
  {(2)} & $n_{0,1,2} \leq n_{0} + n_{1} + n_{3};$ & $n_{3,4}   \leq n_{2} + n_{4};$\\
  {(3)} & $\forall n_{i} \geq 0$ & \\
  \hline
 \end{tabular}
 \caption{{Necessary} constraints of a successful attack.}
 \label{table:security_constraints}
\end{table}

We use an analytical solver~\cite{wolframalpha2019Nov} to {identify} a set of $n_{i}$ values that meets all constraints in Table \ref{table:security_constraints} for the {BlockHammer} configuration we {provide} in Table~\ref{table:tuned_params}.
{We find that there exist{s} no combination of $n_{i}$ values that satisfy these constraints. {Therefore, we conclude} that no access pattern exists that}
can activate an aggressor row more than \nrh{} times within a refresh window {in a BlockHammer-protected system.}

%% file: 06_areaandlatency.tex
\begin{savenotes}
\begin{table*}[ht]
    \centering
    \scriptsize
    \renewcommand{\arraystretch}{0.92} %
    \resizebox{\linewidth}{!}{
    \begin{tabular}{l|rrrrrr|rrrrrr}
        \multirow{3}{*}{\textbf{Mitigation Mechanism}} & \multicolumn{6}{c}{\textbf{\nrh{}=32K*}} & \multicolumn{6}{c}{\textbf{\nrh{}=1K}} \\
            & \textit{SRAM} & \textit{CAM} & \multicolumn{2}{c}{\textit{Area}} & Access Energy & Static Power
            & \textit{SRAM} & \textit{CAM} & \multicolumn{2}{c}{\textit{Area}} & Access Energy & Static Power\\
            & KB & KB & mm$\mathbf{^2}$ & \% CPU & (\SI{}{\pico\joule}) & (\SI{}{\milli\watt}) & KB & KB & mm$\mathbf{^2}$ & \% CPU & (\SI{}{\pico\joule}) & (\SI{}{\milli\watt}) \\\hline
        \textbf{BlockHammer} & \textbf{51.48} & \textbf{1.73} & \textbf{$\mathbf{0.14}$} & \textbf{0.06} & \textbf{20.30} & \textbf{22.27} & \textbf{441.33} & \textbf{55.58} & \textbf{$\mathbf{1.57}$} & \textbf{0.64} & \textbf{99.64} & \textbf{220.99} \\
        \quad Dual counting Bloom filters & 48.00 & - & $0.11$ & 0.04 & 18.11 & 19.81 & 384.00 & - & $0.74$ & 0.30 & 86.29 & 158.46\\
        \quad H3 hash functions & - & - & $<0.01$ & $<0.01$ & - & - & - & - & $<0.01$ & $<0.01$ & - & -\\
        \quad Row activation history buffer & 1.73 & 1.73 & $0.03$ & 0.01 & 1.83 & 2.05 & 55.58 & 55.58 & $0.83$ & 0.34 & 12.99 & 62.12 \\
        \quad AttackThrottler counters & 1.75 & - & $<0.01$ & $<0.01$ & 0.36 & 0.41 & 1.75 & - & $<0.01$ & $<0.01$ & 0.36 & 0.41\\
        \textbf{PARA~\cite{kim2014flipping} }   & \textbf{-} & \textbf{-} & \textbf{$\mathbf{<0.01}$} & \textbf{-} & \textbf{-} & \textbf{-} & \textbf{-} & \textbf{-} & \textbf{$\mathbf{<0.01}$} & \textbf{-} & \textbf{-} & \textbf{-} \\
        \textbf{ProHIT~\cite{son2017making}* } & \textbf{-} & \textbf{0.22} & \textbf{$\mathbf{<0.01}$} & \textbf{<0.01} & \textbf{3.67} & \textbf{0.14} & \textbf{$\times$} & \textbf{$\times$} & \textbf{$\times$} & \textbf{$\times$} & \textbf{$\times$} & \textbf{$\times$}\\
        \textbf{MrLoc~\cite{you2019mrloc}* }  & \textbf{-} & \textbf{0.47} & \textbf{$\mathbf{<0.01}$} & \textbf{<0.01} & \textbf{4.44} & \textbf{0.21} & \textbf{$\times$} & \textbf{$\times$} & \textbf{$\times$} & \textbf{$\times$} & \textbf{$\times$} & \textbf{$\times$}\\
        \textbf{CBT~\cite{seyedzadeh2018cbt} } & \textbf{16.00} & \textbf{8.50} & \textbf{$\mathbf{0.20}$} & \textbf{0.08} & \textbf{9.13} & \textbf{35.55} & \textbf{512.00} & \textbf{272.00} & \textbf{$\mathbf{3.95}$} & \textbf{1.60} & \textbf{127.93} & \textbf{535.50} \\
        \textbf{TWiCE~\cite{lee2019twice} } & \textbf{23.10} & \textbf{14.02} & \textbf{$\mathbf{0.15}$} & \textbf{0.06} & \textbf{7.99} & \textbf{21.28} & \textbf{738.32} & \textbf{448.27} & \textbf{$\mathbf{5.17}$} & \textbf{2.10} & \textbf{124.79} & \textbf{631.98}\\
        \textbf{Graphene~\cite{park2020graphene} } & \textbf{-} & \textbf{5.22} & \textbf{$\mathbf{0.04}$} & \textbf{0.02} & \textbf{40.67} & \textbf{3.11} & \textbf{-} & \textbf{166.03} & \textbf{$\mathbf{1.14}$} & \textbf{0.46} & \textbf{917.55} & \textbf{93.96} \\
        \hline
        \multicolumn{13}{l}{{*~\prohit{}~\cite{son2017making} and \mrloc{}~\cite{you2019mrloc} do {\emph{not}}
        provide a concrete discussion on how to adjust their empirically-determined parameters {for different} \nrh{} {values}. Therefore, we (1)~report their values }}\\
        \multicolumn{13}{l}{{for a fixed design point that each paper provides for \nrh{}=2K and (2)~mark values we cannot estimate using an $\times$.}}
    \end{tabular}
    }
    \caption{{{Per-rank area, {access energy,} and {static} power} of BlockHammer {vs. state-of-the-art RowHammer mitigation mechanisms.}}}
    \label{tab:area_cost_analysis}
\end{table*}
\end{savenotes}
\section{Hardware Complexity Analysis}

{We evaluate BlockHammer's {(1)}~chip area, {static power, and access energy} {consumption} using CACTI~\cite{muralimanohar2009cacti6} and {(2)~circuit latency using} Synopsys DC~\cite{synopsys}. We demonstrate that BlockHammer's physical costs are competitive with state-of-the-art RowHammer mitigation mechanisms.}

\subsection{Area, Static Power, and Access Energy}
\label{sec:blockhammer_areaoverhead}

{\lineskiplimit=-\maxdimen%
Table~\ref{tab:area_cost_analysis} shows an area, {static power, and access energy} cost analysis {of} BlockHammer
alongside six state-of-the-art RowHammer mitigation
mechanisms~\cite{kim2014flipping,son2017making,you2019mrloc,seyedzadeh2018cbt, lee2019twice, park2020graphene}, one of which is concurrent work {with} BlockHammer (Graphene~\cite{park2020graphene}).
{We perform this analysis {at} two RowHammer thresholds {(\nrh{})}: 32K and 1K.\footnote{We {configure each mechanism} as we describe in Section~\ref{sec:methodology}.}}}

\noindent
\textbf{{Main Components of BlockHammer.}} {BlockHammer combines two mechanisms: RowBlocker and AttackThrottler.} RowBlocker, {as shown in Figure~\ref{fig:overview},}
{consists of two components} (1)~{\rbbl{}, which implements} {a dual counting Bloom filter {for each DRAM bank}}, and (2)~{\rbhb{}, which implements} a row activation history buffer {for each DRAM rank}.
When configured {to handle} a RowHammer threshold {(\nrh{})} of 32K, {as shown in Table~\ref{table:tuned_params},} each counting Bloom filter has {1024} {13-bit} counters, stored in an SRAM array. These {counters} are indexed by four H3-class hash functions~\cite{carter1979universal}{, which introduce negligible area overhead (discussed in Section~\ref{sec:mech_detect})}.
\rbhb{}'s history buffer holds {887~entries} per DRAM rank. {E}ach {entry} contains 32~bits for a row ID, a timestamp, and a valid bit.
{AttackThrottler} uses {two counters per thread per DRAM bank}
to measure the \rhli{} of each <thread, bank> pair.
We estimate {BlockHammer's} overall area overhead as \smash{\SI{0.14}{\milli\meter\squared}} per DRAM rank, for a 16-bank DDR4 memory.
{For {a high-end} 28-core {Intel Xeon processor} system with four memory channels and single-rank {DDR4} DIMMs, BlockHammer consumes approximately \smash{\SI{0.55}{\milli\meter\squared}}, which translates to {only} 0.06\% of {the}
{CPU} die area~\cite{wikichipcascade}.}
{When configured for an \nrh{} of 1K, {we reduce} BlockHammer{'s} blacklisting threshold (\nbl{}) from 8K to 512, reducing the CBF counter width from {13~bits to 9~bits}. To avoid false positives {at} the reduced blacklisting threshold, we increase {the} CBF size to 8K. With this modification, BlockHammer's D-CBF consumes \smash{\SI{0.74}{\milli\meter\squared}}. Reducing \nrh{} mandates larger time delays between subsequent row activations targeting a blacklisted row, thereby increasing the history buffer's size from 887 to 27.8K~entries, which translates to \smash{\SI{0.83}{\milli\meter\squared}} chip area. Therefore, BlockHammer's total area overhead {at an \nrh{} of 1K} is \smash{\SI{1.57}{\milli\meter\squared}} or 0.64\% of the {CPU} die area~\cite{wikichipcascade}.}

\noindent
{\textbf{Area Comparison.}
{\graphene{}, \twice{}, and \cbt{} need to store \SI{5.22}{\kilo\byte}, \SI{37.12}{\kilo\byte}, and \SI{24.50}{\kilo\byte}}} of metadata in the memory controller per DRAM {rank, for the same 16-bank DDR4 memory}, which translates to {similarly {low area overheads of}} 0.02\%, 0.06\%, {and} 0.08\% {of the CPU die area}, respectively. {Graphene's area overhead per byte of metadata is larger than other mechanisms because \graphene{} is fully implemented with CAM logic, as shown in Table~\ref{tab:area_cost_analysis}.} {\para{}, \prohit{}, and \mrloc{} are extremely area efficient compared to other mechanisms because they {are probabilistic mechanisms~\cite{kim2014flipping, son2017making, you2019mrloc}}, and thus do not need to store {kilobytes} of metadata to track row activation rates.}

{We repeat our area overhead analysis for future DRAM chips by scaling the RowHammer threshold down to 1K. While BlockHammer {consumes \smash{\SI{1.57}{\milli\meter\squared}} of chip area {to {prevent {bit-flips}} at this lower threshold},}
\twice{}'s {and \cbt{}'s} area overhead {increases} to 3.3x and 2.5x of BlockHammer's.
{We} conclude that BlockHammer scales better than {both \cbt{} and} \twice{} in terms of area overhead.}
{\graphene{}'s} area overhead does not scale as efficiently as BlockHammer with decreasing RowHammer threshold, and becomes comparable to BlockHammer when configured for a RowHammer threshold of 1K.

\noindent
{\textbf{Static Power and Access Energy Comparison.}
When configured for {an} \nrh{} {of} 32K, {BlockHammer consumes \SI{20.30}{\pico\joule} {per access}, which is half of \graphene{}'s access {energy; and} \SI{22.27}{\milli\watt} of static power, which is
{63\% of \cbt{}'s.}}
BlockHammer's {static} power consumption scales more efficiently {than {that of} \cbt{} and \twice{}} {as {\nrh{}} decreases {to 1K}}, {whereas}
\cbt{} and \twice{} {consume 2.42x and 2.86x} {the} static power {of BlockHammer, respectively.}
Similarly, \graphene{}'s access energy and static power drastically increase by 22.56x and 30.2x, respectively, when {\nrh{}} scales down to 1K. As a result, \graphene{} {consumes $9.21\times$ of BlockHammer's {access energy}.}
}

\subsection{Latency Analysis}
\label{sec:blockhammer_latencyoverhead}
We implement BlockHammer in Verilog HDL and synthesize {our design using} Synopsys DC~\cite{synopsys} {with} a \SI{65}{\nano\meter} process technology to evaluate {the} latency impact on memory accesses. According to our RTL model, {which we open source~\cite{blockhammergithub}}, BlockHammer responds {to} {an} \emph{{``{Is this ACT} RowHammer-safe?''}} query (\circled{1} in \figref{overview}) in only \SI{0.97}{\nano\second}. {This} latency {can be hidden because it is} {one-to-}two orders of magnitude smaller than the row access latency (e.g., \SIrange{45}{50}{\nano\second}) that DRAM standards (e.g., DDRx, LPDDRx, GDDRx) enforce~\cite{ghose2019demystifying, jedec2015low, jedec2017}.

%% file: 07_methodology.tex
\section{Experimental Methodology}
\label{sec:methodology}
\label{sec:evaluation_methodology_simulation}
We evaluate BlockHammer's {effect} on a typical DDR4-based memory subsystem's performance and energy consumption {as compared to six prior} RowHammer mitigation mechanisms~\cite{kim2014flipping,son2017making,you2019mrloc,seyedzadeh2018cbt,lee2019twice,park2020graphene}.
We use Ramulator~\cite{Kim2016Ramulator,ramulatorgithub}
for performance evaluation and DRAMPower~\cite{drampower} to estimate {DRAM} energy consumption.
We open-source our infrastructure, which implements both BlockHammer and six state-of-the-art RowHammer mitigation mechanisms~\cite{blockhammergithub}.
Table~\ref{table:system_configuration} shows our system configuration.

 \newcolumntype{C}[1]{>{\let\newline\\\arraybackslash\hspace{0pt}}m{#1}}
 \begin{table}[h]
 \scriptsize
\centering
 \begin{tabular}{l|C{5.8cm}}
 \hline
 \textbf{Processor} & {\SI{3.2}{\giga\hertz}, \{1,8\}~core, 4-wide issue, {128-entry} instr. window}\\ \hline
 \textbf{Last-Level Cache} & {64-byte} cache line, 8-way {set-associative, \SI{16}{\mega\byte}} \\ \hline
 \textbf{Memory Controller} & {64-entry each read and write request queues; Scheduling policy: FR-FCFS~\cite{rixner00, zuravleff1997controller}; Address mapping: MOP~\cite{kaseridis2011minimalistic}} \\ \hline
 \textbf{Main Memory} & DDR4, 1 channel, 1 rank, 4 bank groups, 4 banks/bank group, {64K} rows/bank\\ \hline
 \end{tabular}
 \caption{{Simulated} system configuration.}
 \label{table:system_configuration}
 \end{table}

\noindent
\textbf{Attack Model.}
{W}e compare BlockHammer under the same {RowHammer} attack model ({i.e.,} double-sided attacks \cite{kim2014flipping}) as prior works use~\cite{lee2019twice, seyedzadeh2018cbt, park2020graphene, kim2014flipping, son2017making, you2019mrloc}. {To do so, we halve the RowHammer threshold that BlockHammer uses to account for the cumulative disturbance effect of both aggressor rows (i.e., }
$N_{RH}*=N_{RH}/2$).
In Sections \ref{sec:evaluation_single_core} and \ref{sec:evaluation_multi_core}, we set $N_{RH}*=16K$ (i.e., $N_{RH}=32K$), which is the minimum RowHammer threshold that \twice{}~\cite{lee2019twice} supports~\cite{kim2020revisiting}.
{In Section~\ref{sec:evaluation_tech_scaling}, we} conduct an $N_{RH}$ scaling study for double-sided attacks, {across a} range of $32K > N_{RH} > 1K$, {using parameters provided in Table~\ref{tab:allconfigs}.}

\noindent
\textbf{{Comparison Points}.}
We compare BlockHammer to
a baseline system with no RowHammer {mitigation and to}
{six} {state-of-the-art RowHammer mitigation} mechanisms {that provide RowHammer-safe operation:} {three {are probabilistic mechanisms~\cite{kim2014flipping, son2017making, you2019mrloc} and another three are deterministic mechanisms~\cite{seyedzadeh2018cbt, lee2019twice, park2020graphene}.}}
(1)~\para{}~\cite{kim2014flipping} mitigates RowHammer by injecting an adjacent row activation with a low probability whenever the memory controller {closes a row following an activation}.
{We
tune} \para{}'s probability threshold for a given RowHammer threshold to meet a desired failure probability (we use \smash{$10^{-15}$} as a typical consumer memory reliability target~\cite{cai2012error, cai2017flashtbd, jedec2011failure, luo2016enabling, patel2017reaper}) in a refresh window (\SI{64}{\milli\second}).
(2)~\prohit{}~\cite{son2017making} implements a history table of recent row activations to extend \para{} by {reducing the probability} threshold for more frequently activated rows.
We configure \prohit{} using the default probabilities and parameters provided in \cite{son2017making}.
(3)~\mrloc{}~\cite{you2019mrloc} {extends \para{} by keeping a record of recently-refreshed potential} victim rows in a queue and dynamically {adjusts} the probability threshold, which it uses to decide whether {or} not to refresh the victim row,
based on {the row's temporal} locality information. We implement \mrloc{} by using the empirically-determined parameters {provided in \cite{you2019mrloc}.}
(4) \cbt{}~\cite{seyedzadeh2017cbt} proposes a tree of counters to count the activations for non-uniformly-sized disjoint memory regions, {each} of which is halved in size (i.e., moved to the next level of the tree) every time its activation count reaches a predefined threshold. After being halved a predefined number of times (i.e., after becoming a leaf node in the tree), all rows in the memory region {are} refreshed.
We implement \cbt{} with a six-level tree that contains 125~counters, {and exponentially increase the threshold values across tree levels}
from 1K to the RowHammer threshold (\nrh{}), {as described in \cite{seyedzadeh2018cbt}}.
(5)~\twice{} uses a table of counters to track the activation count of every row. Aiming {for} an area-efficient implementation, \twice{} periodically prunes the activation records of the rows whose activation counts {{cannot reach a {high enough} value to} cause bit-flips}.
{We implement and configure \twice{} for a RowHammer threshold of 32K using the methodology described in the original paper~\cite{lee2019twice}. Unfortunately, \twice{} faces scalability challenges due to time consuming pruning operations, as described in \cite{kim2020revisiting}. To scale \twice{} for smaller RowHammer thresholds, we follow the same methodology as Kim et al.{~\cite{kim2020revisiting}}.}
(6)~\graphene{}~\cite{park2020graphene} {adopts Misra-Gries, a frequent-element detection algorithm~\cite{misra1982finding},  to detect the} most frequently activated rows in a given time window. Graphene maintains a set of counters where it keeps the address and activation count of frequently activated rows. Whenever a row's counter reaches a multiple of a predefined threshold value, \graphene{} refreshes its adjacent rows. We configure \graphene{} by evaluating the equations provided in the original work~\cite{park2020graphene} for a given RowHammer threshold.

\noindent
\textbf{Workloads.}
We evaluate BlockHammer and state-of-the-art RowHammer mitigation mechanisms with {280} {(30~single-core and 250~multiprogrammed)} workloads.
{W}e use {22} memory-intensive benign applications {from the} SPEC CPU2006 benchmark {suite}~\cite{spec2006}, {four disk I/O {applications}
from {the} YCSB benchmark {suite}~\cite{ycsb}, two network I/O {applications} from a commercial network chip~\cite{nxp.networkaccel}, and two synthetic {microbenchmarks}
that mimic non-temporal data copy.}
{We categorize these benign applications based on their row buffer conflicts per kilo instruction ($RBCPKI$) into three categories: {\emph{L}} ($RBCPKI<1$), {\emph{M}} ($1<RBCPKI<5$), and {\emph{H}} ($RBCPKI>5$). \emph{RBCPKI} is {an} indicator of row activation rate, {which is the key workload property that triggers} RowHammer mitigation mechanisms.}
{There are 12, 9, and 9 applications in the {\emph{L}, \emph{M}, and \emph{H}} categories, respectively, {as listed in Table~\ref{tab:workload_list}}}.
{To mimic a double-sided RowHammer attack, we use a synthetic trace that activates two rows {in each bank} as frequently as possible by alternating between them at every row activation (i.e., $R_{A}$, $R_{B}$, $R_{A}$, $R_{B}$, ...).}

We randomly combine these {single-core} {workloads} to create {two types of multiprogrammed workload{s}}: (1)~{125~{workloads}} with \emph{{no RowHammer attack}}, {each} includ{ing} eight benign {threads}; and (2)~{{125~{workloads}} with a \emph{RowHammer attack present}}, {each} includ{ing} one RowHammer {attack and} seven benign threads. {We simulate
each {{multiprogrammed} workload} until each benign thread executes at least 200~million
instructions. For all configurations, we warm up the
caches by fast-forwarding 100~million instructions,} {as done in prior work~\cite{kim2020revisiting}}.

\noindent
\textbf{Performance {and {DRAM} Energy} Metrics.}
{We evaluate BlockHammer's impact on \emph{system throughput} (in terms of weighted speedup~\cite{snavely2000symbiotic, eyerman2008systemlevel, michaud2012demystifying}), \emph{job turnaround time} (in terms of harmonic speedup~\cite{luo2001balancing,eyerman2008systemlevel}), and \emph{fairness} {(in terms of maximum slowdown~\cite{kim2010thread, kim2010atlas,subramanian2014bliss,subramanian2016bliss, subramanian2013mise, mutlu2007stall, subramanian2015application, ebrahimi2012fairness, ebrahimi2011prefetch, das2009application, das2013application}})}.
{Because the performance of a RowHammer attack should not be accounted for in the {performance evaluation},
we calculate all three metrics only for benign {applications}.}
{To evaluate DRAM energy consumption, we compare the total energy consumption that DRAMPower provides in Joules. DRAM energy consumption includes both benign and RowHammer attack requests.}
{Each data point shows the average value across
all workloads, with minimum and maximum values {depicted} {using}
error bars.}

%% file: 08_experimental_evaluation.tex
\section{Performance and Energy Evaluation}
\label{sec:overhead_analysis}
\label{sec:evaluation}

{{W}e evaluate the performance and energy overheads of BlockHammer and {six} state-of-the-art RowHammer mitigation mechanisms.}
{First, we evaluate all mechanisms with single-core {applications} and show that BlockHammer exhibits {no} performance
and energy overheads, compared to a baseline system without any RowHammer mitigation.
Second, we evaluate BlockHammer with multiprogrammed {workloads} and show that, by throttling an attack's requests, BlockHammer significantly improves the performance of benign applications {by 45.4\% on average (with a maximum of 61.9\%),}
compared to
{both {the baseline system} and a system with the prior best-performing state-of-the-art RowHammer mitigation mechanism.}
Third, we compare BlockHammer {with} state-of-the-art RowHammer mitigation mechanisms {when applied to} future DRAM chips {that} are projected to be more vulnerable to RowHammer. {We} show that BlockHammer {is competitive with state-of-the-art mechanisms at RowHammer thresholds as low as 1K when there is no attack in the system, and provides significantly higher performance and lower DRAM energy consumption than state-of-the-art mechanisms when a RowHammer attack is present}.
{Fourth}, we {provide {an} analysis of} BlockHammer's {internal mechanisms.}}

\subsection{Single-Core {Applications}}
\label{sec:evaluation_single_core}
\begin{figure}[h]
    \centering
    \includegraphics[width=\columnwidth]{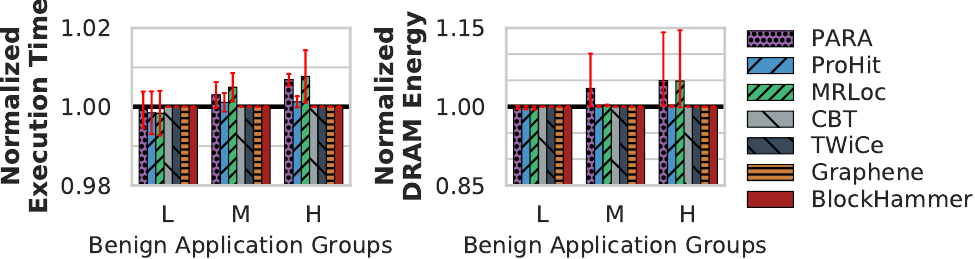}
    \caption{Execution time and DRAM energy consumption for {benign} single-core {applications}, normalized to baseline.}
    \label{fig:single_core_performance_overhead_wo_hammer}
\end{figure}

{\figref{single_core_performance_overhead_wo_hammer} presents the execution time and energy of benign applications (grouped into three categories based on their {\emph{RBCPKI}; see Section~\ref{sec:methodology}}) when executed on a single-core system that uses BlockHammer versus six state-of-the-art mitigation {mechanisms}, normalized to a baseline system that does not employ any RowHammer mitigation mechanism.}

{We observe} that BlockHammer {introduces no} performance {and {DRAM} energy} overheads on benign applications {compared to the baseline configuration}.
This is because benign applications' per-row activation rates {never} exceed BlockHammer's blacklisting threshold (\nbl{}).
{In {contrast},
{\para{}/\mrloc{}} exhibit {0.7\%/0.8\%} performance and {4.9\%/4.9\%} energy overheads for high \emph{RBCPKI} {applications,} on average.} {\cbt{}, \twice{}, and \graphene{} do not perform any victim row refreshes in these {applications} because none of {the applications} activate a row at a high enough rate to trigger victim row refreshes.}
{We} conclude that BlockHammer {does not incur performance or DRAM energy overheads for single-core benign applications.}

\subsection{Multiprogrammed Workload{s}}
\label{sec:evaluation_multi_core}
\begin{figure}[h]
    \centering
    \includegraphics[width=\columnwidth]{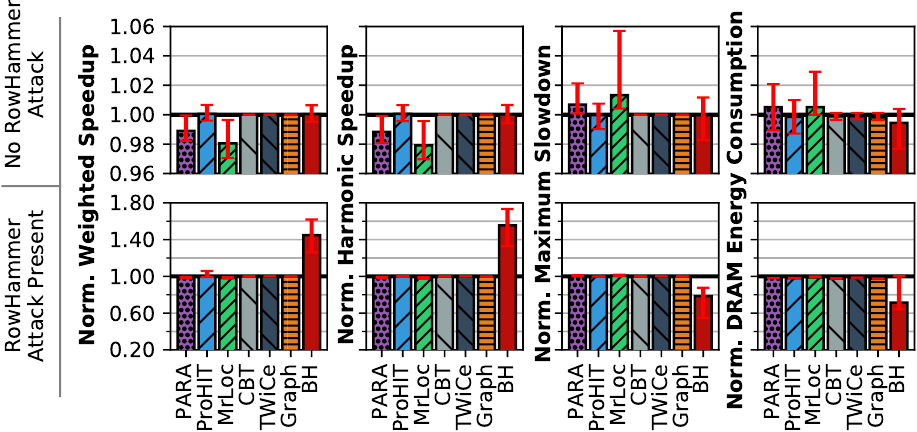}
    \caption{{Performance and {DRAM} energy {consumption} for multiprogrammed {workloads}},
    normalized to {baseline}.}
    \label{fig:8_cores_mech_compare}
    \vspace{0.25\baselineskip}
\end{figure}

{\figref{8_cores_mech_compare} presents the performance and DRAM energy impact of BlockHammer and six state-of-the-art mechanisms\footnote{{We label Graphene as ``Graph'' and BlockHammer as ``BH'' for brevity.}}
on an eight-core system, {normalized to {the} baseline{.}
We show results for two types of workload{s}:
(1)~\emph{No RowHammer Attack}, where all eight applications in the {workload} are benign; and
(2)~\emph{RowHammer Attack Present}, where one of the eight applications in the {workload} is a malicious thread performing a RowHammer attack, running alongside seven benign applications.
We make four observations from the figure.}}

\noindent
{\textbf{No RowHammer Attack.}}
First, BlockHammer {has {a very small} performance overhead}
for multiprogrammed {workloads} {when there is no RowHammer attack present. {BlockHammer incurs} less than 0.5\%, 0.6\%, and 1.2\%
overhead in terms of weighted speedup, harmonic speedup, and maximum slowdown, respectively, compared to the baseline system with no RowHammer mitigation.} In comparison,
{\prohit{}, \cbt{}, \twice{}, and \graphene{} do not perform enough refresh operations to have an impact on system performance, while \para{} and \mrloc{} incur
1.2\% and 2.0\%
performance (i.e., weighted speedup) overheads on average,
respectively}.
{Second,}
BlockHammer \emph{reduces} {average} DRAM energy consumption by
0.6\%,
while for the worst {workload} we observe, it increases energy consumption by up to
0.4\%.
This is {because BlockHammer (1)~increases the standby energy consumption by delaying requests and (2)~reduces} the energy consumed for row activation and precharge operations {by batching delayed requests and servicing them when their target row is activated}.
In comparison, {{\prohit{}}, \cbt{}, \twice{}, and \graphene{} \emph{increase} {average} DRAM energy consumption by less than 0.1\%}, while {\para{} and \mrloc{} \emph{increase} average DRAM energy consumption {by 0.5\%, as a result of}
the unnecessary row refreshes that these mitigation mechanisms must perform.}

\noindent
{\textbf{RowHammer Attack Present.}}
{{{Third}, unlike any other RowHammer mitigation mechanism,}} BlockHammer {\emph{reduces}} the performance degradation
{inflicted on benign applications when one of the applications in the {workload} is a RowHammer attack.}
{By throttling the attack, BlockHammer significantly improves the performance of benign {applications},}
{{with a} 45.0\% (up to 61.9\%) and 56.2\% (up to 73.4\%) increase in weighted and harmonic speedups and 22.7\% (up to 45.4\%) decrease in maximum slowdown on average, respectively.}
{In contrast, \para{}, \prohit{}, and \mrloc{} incur 1.3\%, 0.1\% and {1.7\% performance overheads, on average}, respectively, while {the average performance overheads of} \cbt{}, \twice{}, and \graphene{} are all less than 0.1\%.
{Fourth}, BlockHammer {\emph{reduces}} DRAM energy consumption by {28.9\%} on average (up to 33.8\%). In contrast, {all} other state-of-the-art mechanisms {\emph{increase}} DRAM energy consumption ({by} up to 0.4\%).}
 {BlockHammer} significantly improves performance and DRAM energy because it increases the row buffer locality that benign {applications} experience {by throttling the attacker
 ({the row buffer hit rate increases by 177\% on average, and 23\%} of row buffer conflicts {are converted} {to row buffer misses}).}

We conclude that BlockHammer {(1)~introduces} {very {low}} performance and {DRAM} energy {overheads} for {{workloads} with no RowHammer attack present} and {(2)~significantly} improves
{benign application performance and DRAM energy consumption}
{when a RowHammer attack is present.}

\subsection{{Effect of {Worsening} RowHammer {Vulnerability}}}
\label{sec:evaluation_tech_scaling}

\begin{figure}[ht]
    \centering
    \includegraphics[width=\columnwidth]{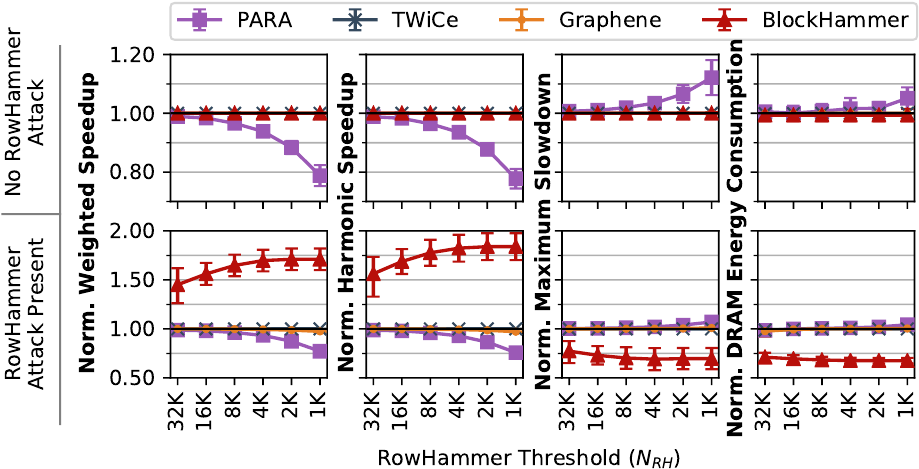}
    \vspace{-5mm}
    \caption{Performance {and DRAM Energy} for different \nrh{} values, normalized to baseline (\nrh{} \emph{decreases} along the x-axis).}
    \label{fig:multi_core_performance_tech_scaling}
    \vspace{2mm}
\end{figure}

{\lineskiplimit=-\maxdimen%
{We analyze {how} BlockHammer's {impact on} performance {and DRAM energy consumption}
{scales as DRAM chips become increasingly  vulnerable to RowHammer (i.e., as the RowHammer threshold, {\nrh{},} decreases)}.}
{We compare BlockHammer with three state-of-the-art RowHammer mitigation mechanisms, {which are shown to be the most viable mechanisms when the RowHammer threshold decreases~\cite{kim2020revisiting, park2020graphene}}: {\para{}~\cite{kim2014flipping}, \twice{}~\cite{lee2019twice},\footnote{{As described in \secref{methodology}, \twice{} faces latency issues, preventing it from scaling when $\nrhn{}<32K$~\cite{kim2020revisiting}. Our scalability analysis assumes a \twice{} variation that solves this issue, the same as \twice{}-Ideal in \cite{kim2020revisiting}}.} and \graphene{}~\cite{park2020graphene}.}}
{We {analyze} the scalability of these mechanisms down to \nrh{}$=1024$, which is approximately an order of magnitude smaller than the minimum observed \nrh{} reported in {current} literature {(i.e., 9600)}~\cite{kim2020revisiting}}.
\figref{multi_core_performance_tech_scaling} shows the performance {and energy overheads of each mechanism for {our} multiprogrammed {workloads}  as \nrh{} decreases, normalized to the baseline system with no RowHammer mitigation}. {We make two observations from {Figure~\ref{fig:multi_core_performance_tech_scaling}}.}}

\noindent
{\textbf{No RowHammer Attack.}}
{First, {{BlockHammer's} performance and {DRAM} energy {consumption}} {are} better
than \para{} {and competitive with other mechanisms} as \nrh{} decreases. When \nrh{}=1024,
{the average performance and {DRAM} energy overheads of BlockHammer, \graphene{}, and \twice{} are
less than 0.6\% because
they do not act aggressively enough to cause significant performance or energy overheads.
On the other hand, \para{} performs reactive refreshes more aggressively with increasing RowHammer vulnerability, which {leads to a} performance overhead {of}
21.2\% and 22.3\% (weighted and harmonic speedup) and an energy overhead of 5.1\% on average.
}

\noindent
{\textbf{RowHammer Attack Present.}}
Second, BlockHammer's {performance and {DRAM} energy benefits increase}
as \nrh{} decreases. At \nrh{}=1024, BlockHammer {more aggressively throttles a RowHammer attack and mitigates {the} performance degradation {of} benign {applications}. As a result, compared to the baseline, BlockHammer improves {average} {performance} by 71.0\% and 83.9\% {(weighted} and harmonic {speedups)} while reducing the maximum slowdown and DRAM energy consumption by 30.4\% and 32.4\%, respectively. {In contrast,} the additional refresh operations that \graphene{} and \twice{} perform {cause} 2.9\% and 0.9\% {average} performance {degradation} {and 0.4\% and 0.2\% DRAM energy increase} for benign application{s,} respectively.} {BlockHammer is the only RowHammer mitigation mechanism that improves performance and energy when a RowHammer attack {is} present in the system.}

{We conclude that (1)~BlockHammer's performance and energy overheads {remain} negligible at reduced RowHammer thresholds {as low as \nrh{}=1K {when there is no RowHammer attack},} and (2)~BlockHammer scalably provides {much} higher performance and lower energy {consumption} {than all} state-of-the-art mechanisms when
{a RowHammer attack is present.}
}
}

\subsection{{Analysis of BlockHammer Internal Mechanisms}}
\label{sec:evaluation_false_positives}
BlockHammer's impact on performance and DRAM energy depends on (1)~the false positive rate of the blacklisting mechanism and (2)~the false positive penalty resulting from delaying row activations.
We calculate (1)~the false positive rate as the number of row activations that are mistakenly delayed by BlockHammer's Bloom filters (i.e., activations to rows that would not have been blacklisted if the filters had no aliasing) as a fraction of all activations,
and (2)~the false positive penalty as the additional time delay a mistakenly-delayed row activation suffers from.
We find that for a configuration where \nrh{}=32K, BlockHammer's false positive rate is 0.010\%, and it increases to only 0.012\% when \nrh{} is scaled down to 1K.
Therefore, BlockHammer successfully avoids delaying more than 99.98\% of benign row activations.
Even though we set \smash{\tdelay{}} to \SI{7.7}{\micro\second}, we observe \SI{1.7}{\micro\second}, \SI{3.9}{\micro\second}, and \SI{7.6}{\micro\second} of delay for the 50th, 90th, and 100th percentile of mistakenly-delayed activations (which are only 0.012\% of all activations).

{Note that the worst-case latency we observe is at least two orders of magnitude smaller than}
typical quality-of-service targets, which are {on} the order of milliseconds~\cite{kasture2016tailbench}.
{Therefore, we} believe that BlockHammer is unlikely to {introduce} quality-of-service violations with {its low} worst-case latency ({on the order of \si{\micro\second}}) and very
low {false positive} rate (0.012\%).

%% file: 09_qualitative_analysis.tex
\section{Comparison of Mitigation Mechanisms}
\label{sec:qualitative_analysis}
We qualitatively compare BlockHammer and a number of published RowHammer mitigation mechanisms, which we classify into four high-level approaches, {as defined in Section~\ref{sec:introduction}}:
($i$)~\emph{increased refresh rate},
($ii$)~\emph{physical isolation},
($iii$)~\emph{reactive refresh},
and ($iv$)~\emph{proactive throttling}.
We evaluate RowHammer mitigation mechanisms across four dimensions: \emph{comprehensive protection}, \emph{compatibility with commodity DRAM chips}, \emph{scaling with RowHammer vulnerability}, and \emph{deterministic protection}.
Table~\ref{table:current_mechanisms} summarizes {our comprehensive} qualitative evaluation.

\input{09_table}

\noindent
\textbf{1. Comprehensive Protection.} A RowHammer mitigation mechanism should comprehensively prevent {\emph{all} potential} RowHammer bit-flips regardless of the methods that an attacker may use to {hammer a DRAM row.}
Unfortunately, {four} key RowHammer mitigation mechanisms~\cite{konoth2018zebram, van2018guardion, brasser2017can, aweke2016anvil}
are effective only against a limited threat model and
have already been defeated by recent attacks~\cite{qiao2016new, gruss2016rowhammer, gruss2018another, cojocar2019eccploit, zhang2019telehammer, kwong2020rambleed}
{because they (1)~trust system components (e.g., hypervisor) that can be used to perform a RowHammer attack~\cite{konoth2018zebram, van2018guardion}; (2)~disregard practical methods (e.g., flipping opcode bits within the attacker's memory space~\cite{brasser2017can}) that can be used to gain root privileges; or (3)~detect RowHammer attacks by relying on hardware performance counters (e.g., LLC miss rate~\cite{aweke2016anvil}), which
can be oblivious to several attack models~\cite{van2016drammer, qiao2016new, gruss2018another, tatar2018throwhammer}.}
In contrast, BlockHammer comprehensively prevents RowHammer bit-flips by monitoring all memory accesses from within the memory controller, even if the entire software stack is compromised and the attacker possesses knowledge about all hardware/software implementation details (e.g., the DRAM chip's RowHammer vulnerability characteristics, BlockHammer's configuration parameters).

\noindent
\textbf{2. Compatibility with Commodity DRAM Chips.} {Especially} given that recent works~\cite{cojocar2020rowhammer, frigo2020trrespass, kim2020revisiting} experimentally observe RowHammer bit-flips on {cutting-edge}
commodity DRAM chips, including ones that are marketed as RowHammer-{free}~\cite{frigo2020trrespass, cojocar2020rowhammer, kim2020revisiting}, it is {important} for a RowHammer mitigation mechanism to be compatible with {\emph{all}} commodity
DRAM chips, current and future. To achieve this, a RowHammer mitigation mechanism should \emph{not} (1)~rely on any proprietary information that DRAM vendors do not share, and (2)~require any modifications to DRAM chip design.
Unfortunately, both physical isolation and reactive refresh mechanisms need to be fully aware of the internal physical layout of DRAM rows or require modifications to DRAM chip design either (1)~to ensure that isolated memory regions are not physically close to each other~\cite{konoth2018zebram, brasser2017can, van2018guardion} or (2)~to identify victim rows that need to be refreshed~\cite{greenfield2012throttling, kim2015architectural, bains2015row, bains2016row, bains2016distributed, aweke2016anvil, kim2014flipping, son2017making, you2019mrloc, seyedzadeh2017cbt, seyedzadeh2018cbt, kang2020cattwo, lee2019twice, park2020graphene}.
In contrast, designing BlockHammer requires knowledge of only six readily-available DRAM parameters:
(1)~\trefw{}: the refresh window,
(2)~\trc{}: the ACT-to-ACT latency,
(3)~\tfaw{}: the four-activation window,
(4)~\nrh{}: the RowHammer threshold,
(5)~the blast radius, and
(6)~the blast impact factor.
Among these parameters, \trefw{}, \trc{}, and \tfaw{} are publicly available in datasheets~\cite{micron2014ddr4, jedec2017, jedec2015low, jedec2015hbm}. \nrh{}, the blast radius, and the blast impact factor
can be obtained from prior characterization works~\cite{kim2014flipping, kim2020revisiting, frigo2020trrespass}.
Therefore, BlockHammer is compatible with all commodity DRAM chips because it does not need any proprietary information about or any modifications to commodity DRAM chips.

\noindent
\textbf{3. Scaling with Increasing RowHammer Vulnerability.}
Since main memory is
a {growing}
system performance and
energy bottleneck~\cite{wulf1995hitting, sites1996stupid, wilkes2001memory, mutlu2021primer,
mutlu2013memory, mutlu2014research, kanev2016profiling,
wang2014bigdatabench, boroumand2018google, deoliveira2021, gomez2021benchmarking},
a RowHammer mitigation mechanism {should exhibit acceptable performance and energy overheads at low area cost when configured for more vulnerable DRAM chips.}

\emph{Increasing the refresh rate}~\cite{kim2014flipping, AppleRefInc} is {\emph{already} a prohibitively expensive} solution {for modern DRAM chips with a} RowHammer threshold of 32K. {This is} because the latency of refreshing rows at a high enough rate to prevent bit-flips overwhelms DRAM's availability, increasing its average performance overhead to 78\%, as shown in \cite{kim2020revisiting}.

\emph{~Physical isolation}~\cite{konoth2018zebram, van2018guardion, brasser2017can} {requires reserving as many rows as {twice} the \emph{blast radius}} {(up to 12 in modern DRAM chips~\cite{kim2020revisiting})} to isolate sensitive data from a potential attacker's memory space. This is expensive for most modern systems where memory capacity is {critical.}
{As} the blast radius {has increased} by 33\%
from 2014~\cite{kim2014flipping} to 2020~\cite{kim2020revisiting}, physical isolation mechanisms can require reserving even more rows when configured for future DRAM chips, {further} reducing the total amount of secure memory available to the system.

{\emph{Reactive refresh} mechanisms~\cite{greenfield2012throttling, kim2015architectural, bains2015row, bains2016row, bains2016distributed, aweke2016anvil, kim2014flipping, son2017making, you2019mrloc, seyedzadeh2017cbt, seyedzadeh2018cbt, kang2020cattwo, lee2019twice, park2020graphene} generally incur increasing performance, energy, and/or area overheads at lower RowHammer thresholds {when configured for more vulnerable DRAM chips.}
{ANVIL} samples hardware performance counters on the order of \SI{}{\milli\second} for a RowHammer threshold (\nrh{}) of 110K~\cite{aweke2016anvil}. However, a RowHammer attack can successfully induce bit-flips in less than \SI{50}{\micro\second} when \nrh{} is reduced to 1K, which significantly increases {ANVIL}'s sampling rate, and thus, its performance {and energy} overheads.
\prohit{} and \mrloc{}~\cite{son2017making, you2019mrloc} do not provide a concrete discussion on how to adjust their empirically-determined parameters, so we cannot demonstrate how their overheads scale as DRAM chips become more vulnerable to RowHammer.
\twice{}~\cite{lee2019twice} faces design challenges to protect DRAM chips when reducing \nrh{} below $32\mathrm{K}$}, as described in Section~\ref{sec:methodology}. Assuming that \twice{} overcomes {its design} challenges (as {also assumed} by prior work~\cite{kim2020revisiting}){,} we scale \twice{} down to \nrh{}$=1\mathrm{K}$ along with three other state-of-the-art mechanisms~\cite{seyedzadeh2018cbt, kim2014flipping, park2020graphene}. Table~\ref{tab:area_cost_analysis} shows that the {CPU die area}, access energy, and static power consumption of \twice{}~\cite{lee2019twice}/\cbt{}~\cite{seyedzadeh2018cbt} drastically increase by {35x/20x,} 15.6x/14.0x, and 29.7x/15.1x, respectively, when \nrh{} is reduced from 32K to 1K.
In contrast, BlockHammer {consumes} only {{30\%/40\%,} 79.8\%/77.8\%, 35\%/41.3\% of \twice{}/\cbt{}'s} {CPU die area,} access energy, and static power, respectively, when configured for \nrh{}$=1\mathrm{K}$.
Section~\ref{sec:evaluation_tech_scaling} shows that \para{}'s average performance and DRAM energy overheads reach 21.2\% and 22.3\%, respectively, when configured for \nrh{}$=1\mathrm{K}$.
{We observe that \graphene{} and BlockHammer are the two most scalable mechanisms with worsening RowHammer vulnerability. When configured for \nrh=1K, BlockHammer (1)~consumes only 11\% of \graphene{}'s access energy (see Table~\ref{tab:area_cost_analysis}) and}
(2)~improves benign applications' performance by 71.0\% and reduces DRAM energy consumption by 32.4\% on average, while \graphene{} incurs 2.9\% {performance and 0.4\% DRAM energy} overheads, as shown in Section~\ref{sec:evaluation_tech_scaling}.

{{Na{\"i}ve} {\emph{proactive throttling}~\cite{greenfield2012throttling, kim2014flipping, mutlu2018rowhammer}} either (1)~blocks all activations targeting a row until the end of the refresh window {once the row's activation count reaches the} RowHammer threshold, or (2)~statically extends each row's activation interval so that no {row's activation count can ever exceed the} RowHammer threshold. The first method has a high area overhead because it requires implementing a counter for each DRAM row~\cite{kim2014flipping, mutlu2018rowhammer}, while
the second method prohibitively increases \trc{}~\cite{jedecddr, jedec2017, jedec2015low, jedec2015hbm}} (e.g., 42.2x/1350.4x for a DRAM chip with \nrh{}=32K/1K)~\cite{kim2014flipping, mutlu2018rowhammer}.
{BlockHammer is the first efficient and scalable proactive throttling-based RowHammer prevention technique.}

\noindent
\textbf{4. Deterministic Prevention.} {To effectively prevent all RowHammer bit-flips, a RowHammer mitigation mechanism should be deterministic, meaning that it should ensure RowHammer-safe operation at all times}
{because} it is important to guarantee zero chance of a security failure for a critical system
whose failure or malfunction may result in severe consequences (e.g., related to loss of lives, environmental damage, or economic loss)~\cite{aven2009identification}.
PARA~\cite{kim2014flipping}, ProHIT~\cite{son2017making}, and MRLoc~\cite{you2019mrloc} are probabilistic by design, and therefore cannot reduce the probability of a successful RowHammer attack to zero like \cbt{}~\cite{seyedzadeh2018cbt}, \twice{}~\cite{lee2019twice}, and \graphene{}~\cite{park2020graphene} {potentially} can.
BlockHammer has the capability to provide zero probability for a successful RowHammer attack by guaranteeing that no row can be activated at a RowHammer-unsafe rate.

%% file: 09_table.tex
\newcommand{\astfootnote}[1]{
\let\oldthefootnote=\thefootnote
\setcounter{footnote}{0}
\renewcommand{\thefootnote}{\fnsymbol{footnote}}
\footnote{#1}
\let\thefootnote=\oldthefootnote
}

\begin{table}[htb]
\centering
\footnotesize
\renewcommand{\arraystretch}{0.895}
\setlength\tabcolsep{5pt}
\begin{tabular}{l|l||c|c|c|c}
 \multicolumn{2}{c||}{}
    & \multirow{8}{*}{\thead{Comprehensive Protection}}
    & \multirow{8}{*}{\thead{Compatible w/ Commodity\\DRAM Chips}}
    & \multirow{8}{*}{\thead{Scaling with RowHammer Vulnerability}}
    & \multirow{8}{*}{\thead{Deterministic Protection}} \\
\multicolumn{2}{c||}{} & & & & \\
\multicolumn{2}{c||}{} & & & & \\
\multicolumn{2}{c||}{} & & & & \\
\multicolumn{2}{c||}{} & & & & \\
\multicolumn{2}{c||}{} & & & & \\
 & & & & & \\
\textbf{Approach} & \textbf{Mechanism} &  & &  &   \\ \hhline{=|=|=|=|=|=}%
\multicolumn{2}{l||}{Increased Refresh Rate~\cite{kim2014flipping, AppleRefInc}} & \cmark  & \cmark  & \xmark & \cmark  \\ \hline
\multirow{3}*{\setlength\tabcolsep{0pt}\begin{tabular}{l}Physical\\Isolation\end{tabular}} & CATT~\cite{brasser2017can}      & \xmark & \xmark & {\xmark} & \cmark  \\ %
 & GuardION~\cite{van2018guardion}    & \xmark & \xmark & {\xmark} & \cmark \\
  & ZebRAM~\cite{konoth2018zebram}   & \xmark & \xmark & {\xmark} & \cmark \\ %
          \hline
          & ANVIL~\cite{aweke2016anvil}     & \xmark & \xmark & {\xmark} & \cmark  \\ %
          & \para{}~\cite{kim2014flipping}  & \cmark & \xmark & {\xmark} & \xmark  \\ %
\multirow{3}*{\setlength\tabcolsep{0pt}\begin{tabular}{l}Reactive\\Refresh\end{tabular}}  & \prohit{}~\cite{son2017making}  & \cmark & \xmark & \xmark & \xmark  \\ %
   & \mrloc{}~\cite{you2019mrloc}    & \cmark & \xmark & \xmark & \xmark \\ %
          & \cbt{}~\cite{seyedzadeh2018cbt} & \cmark & \xmark & {\xmark} & \cmark   \\ %
          & \twice{}~\cite{lee2019twice}    & \cmark & \xmark & {\xmark} & \cmark  \\ %
          & \graphene{}~\cite{park2020graphene} & \cmark & \xmark & {\cmark} & \cmark  \\
          \hline
\multirow{3}*{\setlength\tabcolsep{0pt}\begin{tabular}{l}Proactive\\Throttling\end{tabular}} & Naive Thrott.~\cite{mutlu2018rowhammer} & \cmark & \cmark & \xmark & \cmark  \\ %
& Thrott. Supp.~\cite{greenfield2012throttling}& \cmark & \xmark & \xmark & \cmark  \\ \cline{2-6}
          & \textbf{BlockHammer}                                     & \cmark & \cmark & {\cmark} & \cmark  \\ \hline
\end{tabular}
\caption{{Compari{son of} RowHammer mitigation {mechanisms}.}}
\label{table:current_mechanisms}
\end{table}

%% file: 10_relatedwork.tex
\section{Related Work}
\label{sec:relatedwork}

To our knowledge, BlockHammer is the first work that (1)~prevents RowHammer bit-flips efficiently and scalably without {requiring} any proprietary knowledge of or {modification} to DRAM internals, (2)~satisfies all four of the desired characteristics for a RowHammer mitigation mechanism (as we describe in Section~\ref{sec:qualitative_analysis}), and (3)~improves benign application performance and system energy when the system is under a RowHammer attack.
{Sections~\ref{sec:blockhammer_areaoverhead}, \ref{sec:evaluation}, and \ref{sec:qualitative_analysis} already qualitatively and quantitatively {compare} BlockHammer {to} the most relevant {prior} mechanisms, {demonstrating BlockHammer's benefits}.
This section discusses RowHammer mitigation and memory access throttling works that are loosely related to BlockHammer.}

\noindent
\textbf{In-DRAM Reactive Refresh.} A subset of DRAM standards~\cite{jedec2015low, jedec2017} support a mode called \emph{target row refresh} (TRR), which refreshes rows that are physically nearby an aggressor row without exposing any information about {the in-DRAM} row address mapping outside of DRAM chips. TRRespass~\cite{frigo2020trrespass} demonstrates that existing proprietary implementations of TRR are not sufficient to mitigate RowHammer bit-flips{:} many-sided RowHammer attacks reliably induce and exploit bit-flips in state-of-the-art DRAM chips that already implement {TRR}.

\noindent
\textbf{Making Better DRAM Chips.} A different approach to mitigating RowHammer is to implement
architecture- and device-level techniques that make DRAM chips stronger against RowHammer. CROW~\cite{hassan2019crow} maps potential victim rows into dedicated \emph{copy rows} and mitigates RowHammer bit-flips by serving requests from copy rows. Gomez et al.~\cite{gomez2016dummy} place \emph{dummy cells} in DRAM rows that are engineered to be more susceptible to RowHammer than regular cells, and monitor dummy cell charge levels to detect a RowHammer attack.
{Three other works~\cite{yang2016suppression, ryu2017overcoming, han2021surround} propose manufacturing process enhancements {or {implantation of} additional dopants in transistors to reduce} wordline {crosstalk}.}
{Although} these methods mitigate the RowHammer vulnerability of DRAM chips, they {(1)} cannot be applied to already-deployed commodity DRAM chips {and {(2)} can be high cost} {due to {the} required {extensive} chip modifications.}

\noindent
{\textbf{Other Uses of Throttling.}
Prior works on quality-of-service- and fairness-oriented architectures propose selectively throttling main memory accesses to provide latency guarantees and/or improve fairness across applications}
(e.g., \cite{rixner00, moscibroda2007memory, mutlu2007stall,  mutlu2008parbs, lee2008prefetch, kim2010atlas, kim2010thread,  subramanian2014bliss, subramanian2016bliss, ausavarungnirun2012staged, ebrahimi2011parallel, ebrahimi2012fairness, ebrahimi2011prefetch, nychis2012chip, nychis2010next, chang2012hat, usui2016dash}).
These mechanisms are \emph{not} designed to prevent RowHammer attacks and thus do not interfere with a RowHammer attack when there is no contention between memory accesses.
In contrast, BlockHammer's primary goal is to
{prevent RowHammer attacks from inducing bit-flips.} As such, BlockHammer is {complementary} to these mechanisms, and can work together with them.

%% file: 11_conclusion.tex
\section{Conclusion}
\label{sec:conclusion}

We introduce BlockHammer, a new RowHammer detection and prevention mechanism that
uses area-efficient Bloom filters to track and proactively throttle memory accesses that can potentially induce RowHammer bit-flips. BlockHammer operates entirely from within the memory controller, comprehensively protecting a system from all RowHammer bit-flips at low area, energy, and performance cost. Compared to existing RowHammer mitigation mechanisms, BlockHammer is the first one that (1)~prevents RowHammer bit-flips efficiently and scalably without knowledge of or {modification} to DRAM internals, (2)~provides all four desired characteristics of a RowHammer mitigation mechanism (as we describe in Section~\ref{sec:qualitative_analysis}), and (3)~improves the performance and energy consumption of a system that is under attack. We believe that BlockHammer provides a new direction in RowHammer prevention and hope that it enables researchers and engineers to develop low-cost RowHammer-free systems going forward.
To further aid future research and development, we make BlockHammer's source code freely and openly available~\cite{blockhammergithub}.

%% file: 12_appendix_tables.tex
\section{{Appendix Tables}}
\label{sec:appendix_tables}
\vspace{0.8em}
{Table \ref{tab:allconfigs} shows BlockHammer's configuration parameters used for each RowHammer threshold (\nrh{}) in Sections~\ref{sec:blockhammer_areaoverhead} and~\ref{sec:evaluation_tech_scaling}.}

\begin{table}[h!]
    \centering
    \begin{tabular}{r||rrrr}
        \emph{$\mathbf{\nrhn{}}$} & {\nrhtuned{}} & {CBF Size} & {\nbl{}} & {\tbf{}} \\ \hline\hline
        \textbf{32K} & 16K & 1K & 8K & \SI{64}{\milli\second} \\
        \textbf{16K} &  8K & 1K & 4K & \SI{64}{\milli\second} \\
        \textbf{ 8K} &  4K & 1K & 2K & \SI{64}{\milli\second} \\
        \textbf{ 4K} &  2K & 2K & 1K & \SI{64}{\milli\second} \\
        \textbf{ 2K} &  1K & 4K & 512& \SI{64}{\milli\second} \\
        \textbf{ 1K} & 512 & 8K & 256& \SI{64}{\milli\second} \\
         \hline
    \end{tabular}
    \caption{BlockHammer's configuration parameters used for different $\mathbf{\nrhn{}}$ values.}
    \label{tab:allconfigs}
\end{table}

\vspace{1em}
{Table~\ref{tab:workload_list} {lists the} 30 benign applications we use for cycle-level simulations. We report last-level cache misses ($MPKI$) and row buffer conflicts ($RBCPKI$) per kilo instructions for each application. Non-temporal data copy, YCSB Disk I/O, and network accelerator applications do not have an $MPKI$ value because they directly access main memory.}
\vspace{0.8em}
\begin{table}[h!]
    \centering
    \begin{tabular}{l|l|l||rr}
\textbf{Category}                & \textbf{Benchmark Suite}            & \textbf{Application}    &   \textbf{MPKI} & \textbf{RBCPKI} \\ \hline \hline
\multirow{12}{*}{L}   & \multirow{10}{*}{SPEC2006} & 444.namd       &    0.1 &   0.0  \\
                        &                            & 481.wrf        &    0.1 &   0.0  \\
                        &                            & 435.gromacs    &    0.2 &   0.0  \\
                        &                            & 456.hmmer      &    0.1 &   0.0  \\
                        &                            & 464.h264ref    &    0.1 &   0.0  \\
                        &                            & 447.dealII     &    0.1 &   0.0  \\
                        &                            & 403.gcc        &    0.2 &   0.1  \\
                        &                            & 401.bzip2      &    0.3 &   0.1  \\
                        &                            & 445.gobmk      &    0.4 &   0.1  \\
                        &                            & 458.sjeng      &    0.3 &   0.2  \\ \cline{2-2}
                        & Non-Temp. Data Copy        & movnti.rowmaj  &    -   &   0.2  \\ \cline{2-2}
                        & \multirow{4}{*}{YCSB Disk I/O}      & ycsb.A         &    -   &   0.4  \\ \cline{1-1}\cline{3-5}
\multirow{9}{*}{M} &                            & ycsb.F         &    -   &   1.0  \\
                        &                            & ycsb.C         &    -   &   1.0  \\
                        &                            & ycsb.B         &    -   &   1.1  \\ \cline{2-2}
                        & \multirow{12}{*}{SPEC2006} & 471.omnetpp    &    1.3 &   1.2  \\
                        &                            & 483.xalancbmk  &    8.5 &   2.4  \\
                        &                            & 482.sphinx3    &    9.6 &   3.7  \\
                        &                            & 436.cactusADM  &   16.5 &   3.7  \\
                        &                            & 437.leslie3d   &    9.9 &   4.6  \\
                        &                            & 473.astar      &    5.6 &   4.8  \\ \cline{1-1}\cline{3-5}
\multirow{9}{*}{H}   &                            & 450.soplex     &   10.2 &   7.1  \\
                        &                            & 462.libquantum &   26.9 &   7.7  \\
                        &                            & 433.milc       &   13.6 &  10.9  \\
                        &                            & 459.GemsFDTD   &   20.6 &  15.3  \\
                        &                            & 470.lbm        &   36.5 &  24.7  \\
                        &                            & 429.mcf        &  201.7 &  62.3  \\ \cline{2-2}
                        & Non-Temp. Data Copy        & movnti.colmaj  &  -     &  30.9  \\ \cline{2-2}
                        & \multirow{2}{*}{Network accelerator}   & freescale1     &  -     & 336.8  \\
                        &                            & freescale2     &  -     & 370.4  \\ \hline

    \end{tabular}
    \caption{Benign applications used in cycle-level simulations.}
    \label{tab:workload_list}
\end{table}

%% file: yaglikca2021blockhammer_arxiv.bbl
\begin{thebibliography}{100}
\providecommand{\url}[1]{#1}
\csname url@samestyle\endcsname
\providecommand{\newblock}{\relax}
\providecommand{\bibinfo}[2]{#2}
\providecommand{\BIBentrySTDinterwordspacing}{\spaceskip=0pt\relax}
\providecommand{\BIBentryALTinterwordstretchfactor}{4}
\providecommand{\BIBentryALTinterwordspacing}{\spaceskip=\fontdimen2\font plus
\BIBentryALTinterwordstretchfactor\fontdimen3\font minus
  \fontdimen4\font\relax}
\providecommand{\BIBforeignlanguage}[2]{{%
\expandafter\ifx\csname l@#1\endcsname\relax
\typeout{** WARNING: IEEEtranS.bst: No hyphenation pattern has been}%
\typeout{** loaded for the language `#1'. Using the pattern for}%
\typeout{** the default language instead.}%
\else
\language=\csname l@#1\endcsname
\fi
#2}}
\providecommand{\BIBdecl}{\relax}
\BIBdecl

\bibitem{aga2017good}
M.~T. Aga \emph{et~al.}, ``{When Good Protections Go Bad: Exploiting Anti-DoS
  Measures to Accelerate Rowhammer Attacks},'' in \emph{HOST}, 2017.

\bibitem{AppleRefInc}
{Apple Inc.}, ``{About the Security Content of Mac EFI Security Update
  2015-001},'' \url{https://support.apple.com/en-us/HT204934}, {June 2015}.

\bibitem{ausavarungnirun2012staged}
R.~Ausavarungnirun \emph{et~al.}, ``{Staged Memory Scheduling: Achieving High
  Performance and Scalability in Heterogeneous Systems},'' in \emph{ISCA},
  2012.

\bibitem{aven2009identification}
T.~Aven, ``{Identification of Safety and Security Critical Systems and
  Activities},'' \emph{Reliability Engineering \& System Safety}, 2009.

\bibitem{aweke2016anvil}
Z.~B. Aweke \emph{et~al.}, ``{ANVIL: Software-Based Protection Against
  Next-Generation Rowhammer Attacks},'' in \emph{ASPLOS}, 2016.

\bibitem{bains2015row}
K.~Bains \emph{et~al.}, ``{Row Hammer Refresh Command},'' {U.S.}\ Patent
  9,117,544. 2015.

\bibitem{bains2016distributed}
K.~S. Bains and J.~B. Halbert, ``{Distributed Row Hammer Tracking},'' {U.S.}\
  Patent 9,299,400. 2016.

\bibitem{bains2016row}
K.~S. Bains and J.~B. Halbert, ``{Row Hammer Monitoring Based on Stored Row
  Hammer Threshold Value},'' {U.S.}\ Patent 9,384,821. 2016.

\bibitem{barenghi2018software}
A.~Barenghi \emph{et~al.}, ``{Software-Only Reverse Engineering of Physical
  DRAM Mappings for Rowhammer Attacks},'' in \emph{IVSW}, 2018.

\bibitem{bhattacharya2016curious}
S.~Bhattacharya and D.~Mukhopadhyay, ``{Curious Case of Rowhammer: Flipping
  Secret Exponent Bits Using Timing Analysis},'' in \emph{CHES}, 2016.

\bibitem{bloom1970space}
B.~Bloom, ``{Space/Time Trade-Offs in Hash Coding with Allowable Errors},''
  \emph{CACM}, 1970.

\bibitem{boroumand2018google}
A.~Boroumand \emph{et~al.}, ``{Google Workloads for Consumer Devices:
  Mitigating Data Movement Bottlenecks},'' in \emph{ASPLOS}, 2018.

\bibitem{bosman2016dedup}
E.~Bosman \emph{et~al.}, ``{Dedup Est Machina: Memory Deduplication as An
  Advanced Exploitation Vector},'' in \emph{S\&P}, 2016.

\bibitem{brasser2017can}
F.~Brasser \emph{et~al.}, ``{Can't Touch This: Software-Only Mitigation Against
  Rowhammer Attacks Targeting Kernel Memory},'' in \emph{USENIX Security},
  2017.

\bibitem{cai2017flashtbd}
Y.~Cai \emph{et~al.}, ``{Error Characterization, Mitigation, and Recovery in
  Flash Memory Based Solid-State Drives},'' \emph{Proc. IEEE}, 2017.

\bibitem{cai2012error}
Y.~Cai \emph{et~al.}, ``{Error Patterns in {MLC NAND} Flash Memory:
  Measurement, Characterization, and Analysis},'' in \emph{DATE}, 2012.

\bibitem{carter1979universal}
J.~Carter and M.~Wegman, ``{Universal Classes of Hash Functions},''
  \emph{JCSS}, 1979.

\bibitem{drampower}
K.~Chandrasekar \emph{et~al.}, ``{DRAMPower: Open-Source DRAM Power \& Energy
  Estimation Tool},'' \url{http://www.drampower.info/}.

\bibitem{chang2016understanding}
K.~K. Chang \emph{et~al.}, ``{Understanding Latency Variation in Modern DRAM
  Chips: Experimental Characterization, Analysis, and Optimization},'' in
  \emph{SIGMETRICS}, 2016.

\bibitem{chang2014improving}
K.~K. Chang \emph{et~al.}, ``{Improving DRAM Performance by Parallelizing
  Refreshes with Accesses},'' in \emph{HPCA}, 2014.

\bibitem{chang2016low}
K.~K. Chang \emph{et~al.}, ``{Low-Cost Inter-Linked Subarrays (LISA): Enabling
  Fast Inter-Subarray Data Movement in DRAM},'' in \emph{HPCA}, 2016.

\bibitem{chang2017understanding}
K.~K. Chang \emph{et~al.}, ``{Understanding Reduced-Voltage Operation in Modern
  DRAM Devices: Experimental Characterization, Analysis, and Mechanisms},'' in
  \emph{SIGMETRICS}, 2017.

\bibitem{chang2012hat}
K.~K. Chang \emph{et~al.}, ``{HAT: Heterogeneous Adaptive Throttling for
  On-Chip Networks},'' in \emph{SBAC-PAD}, 2012.

\bibitem{cojocar2020rowhammer}
L.~Cojocar \emph{et~al.}, ``{Are We Susceptible to Rowhammer? An End-to-End
  Methodology for Cloud Providers},'' in \emph{S\&P}, 2020.

\bibitem{cojocar2019eccploit}
L.~Cojocar \emph{et~al.}, ``{Exploiting Correcting Codes: On the Effectiveness
  of ECC Memory Against Rowhammer Attacks},'' in \emph{S\&P}, 2019.

\bibitem{ycsb}
B.~Cooper \emph{et~al.}, ``{Benchmarking Cloud Serving Systems with {YCSB}},''
  in \emph{SoCC}, 2010.

\bibitem{das2013application}
R.~Das \emph{et~al.}, ``{Application-to-Core Mapping Policies to Reduce Memory
  System Interference in Multi-Core Systems},'' in \emph{HPCA}, 2013.

\bibitem{das2009application}
R.~Das \emph{et~al.}, ``{Application-Aware Prioritization Mechanisms for
  On-Chip Networks},'' in \emph{MICRO}, 2009.

\bibitem{ebrahimi2012fairness}
E.~Ebrahimi \emph{et~al.}, ``{Fairness via Source Throttling: A Configurable
  and High Performance Fairness Substrate for Multi Core Memory Systems},'' in
  \emph{ASPLOS}, 2010.

\bibitem{ebrahimi2011prefetch}
E.~Ebrahimi \emph{et~al.}, ``{Prefetch-Aware Shared Resource Management for
  Multi-Core Systems},'' in \emph{ISCA}, 2011.

\bibitem{ebrahimi2011parallel}
E.~Ebrahimi \emph{et~al.}, ``{Parallel Application Memory Scheduling},'' in
  \emph{MICRO}, 2011.

\bibitem{eyerman2008systemlevel}
S.~Eyerman and L.~Eeckhout, ``{System-Level Performance Metrics for
  Multiprogram Workloads},'' \emph{IEEE Micro}, 2008.

\bibitem{fan2000summary}
L.~Fan \emph{et~al.}, ``{Summary Cache: A Scalable Wide-Area Web Cache Sharing
  Protocol},'' \emph{TON}, 2000.

\bibitem{frigo2018grand}
P.~Frigo \emph{et~al.}, ``{Grand Pwning Unit: Accelerating Microarchitectural
  Attacks with the GPU},'' in \emph{S\&P}, 2018.

\bibitem{frigo2020trrespass}
P.~Frigo \emph{et~al.}, ``{TRRespass: Exploiting the Many Sides of Target Row
  Refresh},'' in \emph{{S\&P}}, 2020.

\bibitem{ghose2019demystifying}
S.~Ghose \emph{et~al.}, ``{Demystifying Complex Workload--DRAM Interactions: An
  Experimental Study},'' in \emph{{SIGMETRICS}}, 2019.

\bibitem{vampire2018ghose}
S.~Ghose \emph{et~al.}, ``{What Your DRAM Power Models Are Not Telling You:
  Lessons from a Detailed Experimental Study},'' in \emph{SIGMETRICS}, 2018.

\bibitem{gomez2016dummy}
H.~{Gomez} \emph{et~al.}, ``{{DRAM} Row-Hammer Attack Reduction Using Dummy
  Cells},'' in \emph{NORCAS}, 2016.

\bibitem{gomez2021benchmarking}
J.~G{\'o}mez-Luna \emph{et~al.}, ``{Benchmarking a New Paradigm: Understanding
  a Modern Processing-in-Memory Architecture},'' in \emph{{SIGMETRICS}}, 2021.

\bibitem{greenfield2012throttling}
Z.~Greenfield and T.~Levy, ``{Throttling Support for Row-Hammer Counters},''
  {U.S.\ Patent 9,251,885}. 2016.

\bibitem{gruss2018another}
D.~Gruss \emph{et~al.}, ``{Another Flip in the Wall of Rowhammer Defenses},''
  in \emph{S\&P}, 2018.

\bibitem{gruss2016rowhammer}
D.~Gruss \emph{et~al.}, ``{Rowhammer.js: A Remote Software-Induced Fault Attack
  in Javascript},'' arXiv:1507.06955 [cs.CR], 2016.

\bibitem{han2021surround}
J.~Han \emph{et~al.}, ``{Surround Gate Transistor With Epitaxially Grown Si
  Pillar and Simulation Study on Soft Error and Rowhammer Tolerance for
  DRAM},'' \emph{TED}, 2021.

\bibitem{hassan2019crow}
H.~{Hassan} \emph{et~al.}, ``{CROW: A Low-Cost Substrate for Improving DRAM
  Performance, Energy Efficiency, and Reliability},'' in \emph{ISCA}, 2019.

\bibitem{hassan2016chargecache}
H.~Hassan \emph{et~al.}, ``{ChargeCache: Reducing DRAM Latency by Exploiting
  Row Access Locality},'' in \emph{HPCA}, 2016.

\bibitem{hassan2017softmc}
H.~Hassan \emph{et~al.}, ``{SoftMC: A Flexible and Practical Open-Source
  Infrastructure for Enabling Experimental DRAM Studies},'' in \emph{HPCA},
  2017.

\bibitem{hong2019terminal}
S.~Hong \emph{et~al.}, ``{Terminal Brain Damage: Exposing the Graceless
  Degradation in Deep Neural Networks Under Hardware Fault Attacks},'' in
  \emph{USENIX Security}, 2019.

\bibitem{horiguchi1997redundancy}
M.~Horiguchi, ``{Redundancy Techniques for High-Density DRAMs},'' in
  \emph{ISIS}, 1997.

\bibitem{itoh2013vlsi}
K.~Itoh, \emph{{VLSI Memory Chip Design}}.\hskip 1em plus 0.5em minus
  0.4em\relax Springer, 2001.

\bibitem{jang2017sgx}
Y.~Jang \emph{et~al.}, ``{SGX-Bomb: Locking Down the Processor via Rowhammer
  Attack},'' in \emph{SOSP}, 2017.

\bibitem{jedecddr}
{{JEDEC}}, \emph{{{{JESD}79F: Double Data Rate (DDR) SDRAM Standard}}}, 2008.

\bibitem{jedec2011failure}
{JEDEC}, \emph{{JEP122G: Failure Mechanisms and Models for Semiconductor
  Devices}}, 2012.

\bibitem{jedec2015low}
{JEDEC}, \emph{{JESD209-4B: Low Power Double Data Rate 4 (LPDDR4) Standard}},
  2017.

\bibitem{jedec2015hbm}
{JEDEC}, \emph{{JESD235C: High Bandwidth Memory (HBM) DRAM}}, 2020.

\bibitem{jedec2017}
{JEDEC}, \emph{{JESD79-4C: DDR4 SDRAM Standard}}, 2020.

\bibitem{ji2019pinpoint}
S.~Ji \emph{et~al.}, ``{Pinpoint Rowhammer: Suppressing Unwanted Bit Flips on
  Rowhammer Attacks},'' in \emph{ASIACCS}, 2019.

\bibitem{kandemir2015memory}
M.~Kandemir \emph{et~al.}, ``{Memory Row Reuse Distance and Its Role in
  Optimizing Application Performance},'' in \emph{SIGMETRICS}, 2015.

\bibitem{kanev2016profiling}
S.~Kanev \emph{et~al.}, ``{Profiling a Warehouse-Scale Computer},'' in
  \emph{ISCA}, 2015.

\bibitem{kang2020cattwo}
I.~Kang \emph{et~al.}, ``{CAT-TWO: Counter-Based Adaptive Tree, Time Window
  Optimized for {DRAM} Row-Hammer Prevention},'' \emph{{IEEE} Access}, 2020.

\bibitem{kaseridis2011minimalistic}
D.~Kaseridis \emph{et~al.}, ``{Minimalist Open-Page: A DRAM Page-Mode
  Scheduling Policy for the Many-Core Era},'' in \emph{MICRO}, 2011.

\bibitem{kasture2016tailbench}
H.~Kasture and D.~Sanchez, ``{TailBench: A Benchmark Suite and Evaluation
  Methodology for Latency-Critical Applications},'' in \emph{IISWC}, 2016.

\bibitem{keeth2001dram}
B.~Keeth and R.~Baker, \emph{{DRAM Circuit Design: A Tutorial}}.\hskip 1em plus
  0.5em minus 0.4em\relax Wiley, 2001.

\bibitem{khan2018sttramrowhammer}
M.~N.~I. {Khan} and S.~{Ghosh}, ``{Analysis of Row Hammer Attack on STTRAM},''
  in \emph{ICCD}, 2018.

\bibitem{khan2014efficacy}
S.~Khan \emph{et~al.}, ``{The Efficacy of Error Mitigation Techniques for DRAM
  Retention Failures: A Comparative Experimental Study},'' in
  \emph{SIGMETRICS}, 2014.

\bibitem{khan2016parbor}
S.~Khan \emph{et~al.}, ``{PARBOR: An Efficient System-Level Technique to Detect
  Data-Dependent Failures in DRAM},'' in \emph{DSN}, 2016.

\bibitem{khan2016case}
S.~Khan \emph{et~al.}, ``{A Case for Memory Content-Based Detection and
  Mitigation of Data-Dependent Failures in DRAM},'' \emph{CAL}, 2016.

\bibitem{khan2017detecting}
S.~Khan \emph{et~al.}, ``{Detecting and Mitigating Data-Dependent DRAM Failures
  by Exploiting Current Memory Content},'' in \emph{MICRO}, 2017.

\bibitem{kim2015architectural}
D.-H. Kim \emph{et~al.}, ``{Architectural Support for Mitigating Row Hammering
  in DRAM Memories},'' \emph{CAL}, 2015.

\bibitem{kim2018solar}
J.~S. Kim \emph{et~al.}, ``{Solar-DRAM: Reducing DRAM Access Latency by
  Exploiting the Variation in Local Bitlines},'' in \emph{ICCD}, 2018.

\bibitem{kim2018dram}
J.~S. Kim \emph{et~al.}, ``{The DRAM Latency PUF: Quickly Evaluating Physical
  Unclonable Functions by Exploiting the Latency--Reliability Tradeoff in
  Modern Commodity DRAM Devices},'' in \emph{HPCA}, 2018.

\bibitem{kim2019d}
J.~S. Kim \emph{et~al.}, ``{D-RaNGe: Using Commodity DRAM Devices to Generate
  True Random Numbers with Low Latency and High Throughput},'' in \emph{HPCA},
  2019.

\bibitem{kim2020revisiting}
J.~S. Kim \emph{et~al.}, ``{Revisiting RowHammer: An Experimental Analysis of
  Modern Devices and Mitigation Techniques},'' in \emph{ISCA}, 2020.

\bibitem{kim2014flipping}
Y.~{Kim} \emph{et~al.}, ``{Flipping Bits in Memory Without Accessing Them: An
  Experimental Study of DRAM Disturbance Errors},'' in \emph{ISCA}, 2014.

\bibitem{kim2010atlas}
Y.~Kim \emph{et~al.}, ``{ATLAS: A Scalable and High-Performance Scheduling
  Algorithm for Multiple Memory Controllers},'' in \emph{HPCA}, 2010.

\bibitem{kim2010thread}
Y.~Kim \emph{et~al.}, ``{Thread Cluster Memory Scheduling: Exploiting
  Differences in Memory Access Behavior},'' in \emph{MICRO}, 2010.

\bibitem{salp}
Y.~Kim \emph{et~al.}, ``{A Case for Exploiting Subarray-Level Parallelism
  (SALP) in DRAM},'' in \emph{ISCA}, 2012.

\bibitem{Kim2016Ramulator}
Y.~Kim \emph{et~al.}, ``{Ramulator: A Fast and Extensible DRAM Simulator},''
  \emph{CAL}, 2016.

\bibitem{konoth2018zebram}
R.~K. Konoth \emph{et~al.}, ``{ZebRAM: Comprehensive and Compatible Software
  Protection Against Rowhammer Attacks},'' in \emph{OSDI}, 2018.

\bibitem{kwong2020rambleed}
A.~Kwong \emph{et~al.}, ``{RAMBleed: Reading Bits in Memory Without Accessing
  Them},'' in \emph{S\&P}, 2020.

\bibitem{lee2008prefetch}
C.~J. Lee \emph{et~al.}, ``{Prefetch-Aware DRAM Controllers},'' in
  \emph{MICRO}, 2008.

\bibitem{lee2017design}
D.~Lee \emph{et~al.}, ``{Design-Induced Latency Variation in Modern DRAM Chips:
  Characterization, Analysis, and Latency Reduction Mechanisms},'' in
  \emph{SIGMETRICS}, 2017.

\bibitem{lee2013tiered}
D.~Lee \emph{et~al.}, ``{Tiered-Latency DRAM: A Low Latency and Low Cost DRAM
  Architecture},'' in \emph{HPCA}, 2013.

\bibitem{lee2015decoupled}
D.~Lee \emph{et~al.}, ``{Decoupled Direct Memory Access: Isolating CPU and IO
  Traffic by Leveraging a Dual-Data-Port DRAM},'' in \emph{PACT}, 2015.

\bibitem{lee2019twice}
E.~Lee \emph{et~al.}, ``{TWiCe: Preventing Row-Hammering by Exploiting Time
  Window Counters},'' in \emph{ISCA}, 2019.

\bibitem{lee2014green}
J.~Lee, ``{Green Memory Solution},'' {Investor’s Forum}, {Samsung
  Electronics}, 2014.

\bibitem{li2012compression}
Z.~Li \emph{et~al.}, ``{Compression of Pending Interest Table with Bloom Filter
  in Content Centric Network},'' in \emph{CFI}, 2012.

\bibitem{lipp2018nethammer}
M.~Lipp \emph{et~al.}, ``{Nethammer: Inducing Rowhammer Faults Through Network
  Requests},'' arXiv:1805.04956 [cs.CR], 2018.

\bibitem{liu2013experimental}
J.~Liu \emph{et~al.}, ``{An Experimental Study of Data Retention Behavior in
  Modern DRAM Devices},'' in \emph{ISCA}, 2013.

\bibitem{liu2012raidr}
J.~Liu \emph{et~al.}, ``{RAIDR: Retention-Aware Intelligent DRAM Refresh},'' in
  \emph{ISCA}, 2012.

\bibitem{luo2020clrdram}
H.~Luo \emph{et~al.}, ``{CLR-DRAM: A Low-Cost DRAM Architecture Enabling
  Dynamic Capacity-Latency Trade-Off},'' in \emph{{ISCA}}, 2020.

\bibitem{luo2001balancing}
K.~Luo \emph{et~al.}, ``{Balancing Thoughput and Fairness in SMT Processors},''
  in \emph{ISPASS}, 2001.

\bibitem{luo2016enabling}
Y.~Luo \emph{et~al.}, ``{Enabling Accurate and Practical Online Flash Channel
  Modeling for Modern {MLC NAND} Flash Memory},'' in \emph{JSAC}, 2016.

\bibitem{meza2015revisiting}
J.~Meza \emph{et~al.}, ``{Revisiting Memory Errors in Large-Scale Production
  Data Centers: Analysis and Modeling of New Trends from the Field},'' in
  \emph{{DSN}}, 2015.

\bibitem{michaud2012demystifying}
P.~Michaud, ``{Demystifying Multicore Throughput Metrics},'' \emph{{CAL}},
  2012.

\bibitem{micron2014ddr4}
{Micron Technology}, ``{SDRAM, 4Gb: x4, x8, x16 DDR4 SDRAM Features},'' 2014.

\bibitem{micron2014networking}
{Micron Technology}, ``{TN-40-03: DDR4 Networking Design Guide},'' 2014.

\bibitem{misra1982finding}
J.~Misra and D.~Gries, ``{Finding Repeated Elements},'' \emph{{Science of
  Computer Programming}}, 1982.

\bibitem{moscibroda2007memory}
T.~Moscibroda and O.~Mutlu, ``{Memory Performance Attacks: Denial of Memory
  Service in Multi-Core Systems},'' in \emph{USENIX Security}, 2007.

\bibitem{muralimanohar2009cacti6}
N.~Muralimanohar \emph{et~al.}, ``{CACTI 6.0: A Tool to Model Large Caches},''
  HP Laboratories, Tech. Rep. HPL-2009-85, 2009.

\bibitem{mutlu2013memory}
O.~Mutlu, ``{Memory Scaling: A Systems Architecture Perspective},'' in
  \emph{IMW}, 2013.

\bibitem{mutlu2017rowhammer}
O.~Mutlu, ``{The RowHammer Problem and Other Issues We May Face as Memory
  Becomes Denser},'' in \emph{DATE}, 2017.

\bibitem{mutlu2018rowhammer}
O.~Mutlu, ``{RowHammer},''
  \url{https://people.inf.ethz.ch/omutlu/pub/onur-Rowhammer-TopPicksinHardwareEmbeddedSecurity-November-8-2018.pdf},
  {Top Picks in Hardware and Embedded Security}. 2018.

\bibitem{mutlu2021primer}
O.~Mutlu \emph{et~al.}, ``{A Modern Primer on Processing in Memory},'' in
  \emph{{arXiv}}, 2020.

\bibitem{mutlu2019rowhammer}
O.~Mutlu and J.~S. Kim, ``{RowHammer: A Retrospective},'' \emph{TCAD}, 2019.

\bibitem{mutlu2007stall}
O.~Mutlu and T.~Moscibroda, ``{Stall-Time Fair Memory Access Scheduling for
  Chip Multiprocessors},'' in \emph{MICRO}, 2007.

\bibitem{mutlu2008parbs}
O.~Mutlu and T.~Moscibroda, ``{Parallelism-Aware Batch Scheduling: Enhancing
  Both Performance and Fairness of Shared DRAM Systems},'' in \emph{ISCA},
  2008.

\bibitem{mutlu2014research}
O.~Mutlu and L.~Subramanian, ``{Research Problems and Opportunities in Memory
  Systems},'' \emph{SUPERFRI}, 2014.

\bibitem{nxp.networkaccel}
{NXP Semiconductors}, ``{QorIQ Processing Platforms: 64-Bit Multicore SoCs},''
  \url{https://www.nxp.com/products/processors-and-microcontrollers/applications-processors/qoriq-platforms:QORIQ_HOME}.

\bibitem{nychis2010next}
G.~Nychis \emph{et~al.}, ``{Next Generation On-Chip Networks: What Kind of
  Congestion Control Do We Need?}'' in \emph{HOTNETS}, 2010.

\bibitem{nychis2012chip}
G.~Nychis \emph{et~al.}, ``{On-Chip Networks from a Networking Perspective:
  Congestion and Scalability in Many-Core Interconnects},'' in \emph{SIGCOMM},
  2012.

\bibitem{deoliveira2021}
G.~F. Oliveira \emph{et~al.}, ``{A New Methodology and Open-Source Benchmark
  Suite for Evaluating Data Movement Bottlenecks: A Near-Data Processing Case
  Study},'' in \emph{SIGMETRICS}, 2021.

\bibitem{park2016statistical}
K.~Park \emph{et~al.}, ``{Statistical Distributions of Row-Hammering Induced
  Failures in DDR3 Components},'' \emph{Microelectronics Reliability}, 2016.

\bibitem{park2020graphene}
Y.~Park \emph{et~al.}, ``{Graphene: Strong yet Lightweight Row Hammer
  Protection},'' in \emph{MICRO}, 2020.

\bibitem{patel2020beer}
M.~Patel \emph{et~al.}, ``{Bit-Exact ECC Recovery (BEER): Determining DRAM
  On-Die ECC Functions by Exploiting DRAM Data Retention Characteristics},'' in
  \emph{{MICRO}}, 2020.

\bibitem{patel2019understanding}
M.~Patel \emph{et~al.}, ``{Understanding and Modeling On-Die Error Correction
  in Modern DRAM: An Experimental Study Using Real Devices},'' in \emph{DSN},
  2019.

\bibitem{patel2017reaper}
M.~Patel \emph{et~al.}, ``{The Reach Profiler (REAPER): Enabling the Mitigation
  of DRAM Retention Failures via Profiling at Aggressive Conditions},'' in
  \emph{ISCA}, 2017.

\bibitem{pessl2016drama}
P.~Pessl \emph{et~al.}, ``{DRAMA: Exploiting DRAM Addressing for Cross-CPU
  Attacks},'' in \emph{USENIX Security}, 2016.

\bibitem{qiao2016new}
R.~Qiao and M.~Seaborn, ``{A New Approach for RowHammer Attacks},'' in
  \emph{HOST}, 2016.

\bibitem{qureshi2015avatar}
M.~Qureshi \emph{et~al.}, ``{AVATAR: A Variable-Retention-Time (VRT) Aware
  Refresh for DRAM Systems},'' in \emph{DSN}, 2015.

\bibitem{razavi2016flip}
K.~Razavi \emph{et~al.}, ``{Flip Feng Shui: Hammering a Needle in the Software
  Stack},'' in \emph{USENIX Security}, 2016.

\bibitem{redeker2002investigation}
M.~Redeker \emph{et~al.}, ``{An Investigation into Crosstalk Noise in DRAM
  Structures},'' in \emph{MTDT}, 2002.

\bibitem{rixner00}
S.~Rixner \emph{et~al.}, ``{Memory Access Scheduling},'' in \emph{ISCA}, 2000.

\bibitem{ryu2017overcoming}
S.-W. Ryu \emph{et~al.}, ``{Overcoming the Reliability Limitation in the
  Ultimately Scaled DRAM using Silicon Migration Technique by Hydrogen
  Annealing},'' in \emph{IEDM}, 2017.

\bibitem{blockhammergithub}
{SAFARI Research Group}, ``{BlockHammer --- GitHub Repository},''
  \url{https://github.com/CMU-SAFARI/blockhammer}.

\bibitem{ramulatorgithub}
{SAFARI Research Group}, ``{Ramulator --- GitHub Repository},''
  \url{https://github.com/CMU-SAFARI/ramulator}.

\bibitem{rowhammergithub}
{SAFARI Research Group}, ``{RowHammer --- GitHub Repository},''
  \url{https://github.com/CMU-SAFARI/rowhammer}.

\bibitem{seaborn2015exploiting}
M.~Seaborn and T.~Dullien, ``{Exploiting the DRAM Rowhammer Bug to Gain Kernel
  Privileges},'' \emph{Black Hat}, 2015.

\bibitem{seshadri2013rowclone}
V.~Seshadri \emph{et~al.}, ``{RowClone: Fast and Energy-Efficient In-DRAM Bulk
  Data Copy and Initialization},'' in \emph{MICRO}, 2013.

\bibitem{seshadri2017ambit}
V.~Seshadri \emph{et~al.}, ``{Ambit: In-Memory Accelerator for Bulk Bitwise
  Operations Using Commodity DRAM Technology},'' in \emph{MICRO}, 2017.

\bibitem{seshadri2015gather}
V.~Seshadri \emph{et~al.}, ``{Gather-Scatter DRAM: In-DRAM Address Translation
  to Improve the Spatial Locality of Non-Unit Strided Accesses},'' in
  \emph{MICRO}, 2015.

\bibitem{seshadri2019dram}
V.~Seshadri and O.~Mutlu, ``{In-DRAM Bulk Bitwise Execution Engine},''
  \emph{arXiv:1905.09822}, 2019.

\bibitem{seyedzadeh2018cbt}
S.~M. {Seyedzadeh} \emph{et~al.}, ``{Mitigating Wordline Crosstalk Using
  Adaptive Trees of Counters},'' in \emph{ISCA}, 2018.

\bibitem{seyedzadeh2017cbt}
S.~M. Seyedzadeh \emph{et~al.}, ``{Counter-Based Tree Structure for Row
  Hammering Mitigation in DRAM},'' \emph{CAL}, 2017.

\bibitem{sites1996stupid}
R.~Sites, ``{It's the Memory, Stupid},'' \emph{Microprocessor Report}, 1996.

\bibitem{smith1981laser}
R.~T. Smith \emph{et~al.}, ``{Laser Programmable Redundancy and Yield
  Improvement in a 64K DRAM},'' \emph{JSSC}, 1981.

\bibitem{snavely2000symbiotic}
A.~Snavely and D.~M. Tullsen, ``{Symbiotic Jobscheduling for A Simultaneous
  Multithreaded Processor},'' in \emph{{ASPLOS}}, 2000.

\bibitem{son2017making}
M.~Son \emph{et~al.}, ``{Making DRAM Stronger Against Row Hammering},'' in
  \emph{DAC}, 2017.

\bibitem{spec2006}
{Standard Performance Evaluation Corp.}, ``{SPEC CPU 2006},''
  \url{http://www.spec.org/cpu2006/}.

\bibitem{subramanian2014bliss}
L.~Subramanian \emph{et~al.}, ``{The Blacklisting Memory Scheduler: Achieving
  High Performance and Fairness at Low Cost},'' in \emph{ICCD}, 2014.

\bibitem{subramanian2016bliss}
L.~Subramanian \emph{et~al.}, ``{BLISS: Balancing Performance, Fairness and
  Complexity in Memory Access Scheduling},'' \emph{TPDS}, 2016.

\bibitem{subramanian2015application}
L.~Subramanian \emph{et~al.}, ``{The Application Slowdown Model: Quantifying
  and Controlling the Impact of Inter-Application Interference at Shared Caches
  and Main Memory},'' in \emph{MICRO}, 2015.

\bibitem{subramanian2013mise}
L.~Subramanian \emph{et~al.}, ``{MISE: Providing Performance Predictability and
  Improving Fairness in Shared Main Memory Systems},'' in \emph{HPCA}, 2013.

\bibitem{synopsys}
{Synopsys, Inc.}, ``{Synopsys Design Compiler},''
  \url{https://www.synopsys.com/support/training/rtl-synthesis/design-compiler-rtl-synthesis.html}.

\bibitem{tatar2018defeating}
A.~Tatar \emph{et~al.}, ``{Defeating Software Mitigations Against Rowhammer: A
  Surgical Precision Hammer},'' in \emph{RAID}, 2018.

\bibitem{tatar2018throwhammer}
A.~Tatar \emph{et~al.}, ``{Throwhammer: {Rowhammer} {Attacks} Over the
  {Network} and {Defenses}},'' in \emph{{USENIX} {ATC}}, 2018.

\bibitem{usui2016dash}
H.~Usui \emph{et~al.}, ``{DASH: Deadline-Aware High-Performance Memory
  Scheduler for Heterogeneous Systems with Hardware Accelerators},''
  \emph{TACO}, 2016.

\bibitem{van2016drammer}
V.~van~der Veen \emph{et~al.}, ``{Drammer: Deterministic Rowhammer Attacks on
  Mobile Platforms},'' in \emph{CCS}, 2016.

\bibitem{van2018guardion}
V.~van~der Veen \emph{et~al.}, ``{GuardION: Practical Mitigation of DMA-Based
  Rowhammer Attacks on ARM},'' in \emph{{DIMVA}}, 2018.

\bibitem{wang2014bigdatabench}
L.~Wang \emph{et~al.}, ``{Bigdatabench: A Big Data Benchmark Suite from
  Internet Services},'' in \emph{HPCA}, 2014.

\bibitem{wang2020figaro}
Y.~Wang \emph{et~al.}, ``{FIGARO: Improving System Performance via Fine-Grained
  In-DRAM Data Relocation and Caching},'' in \emph{MICRO}, 2020.

\bibitem{weissman2020jackhammer}
Z.~Weissman \emph{et~al.}, ``{JackHammer: Efficient Rowhammer on Heterogeneous
  FPGA--CPU Platforms},'' arXiv:1912.11523 [cs.CR], 2020.

\bibitem{wikichipcascade}
WikiChip, ``{Cascade Lake SP - Intel},''
  \url{https://en.wikichip.org/wiki/intel/cores/cascade\_lake\_sp}.

\bibitem{wilkes2001memory}
M.~V. Wilkes, ``{The Memory Gap and the Future of High Performance Memories},''
  \emph{CAN}, 2001.

\bibitem{wolframalpha2019Nov}
{Wolfram Research, Inc.}, ``{WolframAlpha},''
  \url{http://www.wolframalpha.com/}.

\bibitem{wulf1995hitting}
W.~A. Wulf and S.~A. McKee, ``{Hitting the Memory Wall: Implications of the
  Obvious},'' \emph{CAN}, 1995.

\bibitem{xiao2016one}
Y.~Xiao \emph{et~al.}, ``{One Bit Flips, One Cloud Flops: Cross-VM Row Hammer
  Attacks and Privilege Escalation},'' in \emph{USENIX Security}, 2016.

\bibitem{yaglikci2020blockhammerarxiv}
A.~G. Ya{\u{g}}l{\i}k{\c{c}}{\i} \emph{et~al.}, ``{BlockHammer: Preventing
  RowHammer at Low Cost by Blacklisting Rapidly-Accessed DRAM Rows},''
  \emph{arXiv}, 2021.

\bibitem{yang2016suppression}
C.~Yang \emph{et~al.}, ``{Suppression of RowHammer Effect by Doping Profile
  Modification in Saddle-Fin Array Devices for Sub-30-nm DRAM Technology},''
  \emph{TDMR}, 2016.

\bibitem{yang2019trap}
T.~Yang and X.-W. Lin, ``{Trap-Assisted DRAM Row Hammer Effect},'' \emph{EDL},
  2019.

\bibitem{yao2020deephammer}
F.~Yao \emph{et~al.}, ``{Deephammer: Depleting the Intelligence of Deep Neural
  Networks Through Targeted Chain of Bit Flips},'' in \emph{USENIX Security},
  2020.

\bibitem{you2019mrloc}
J.~M. You and J.-S. Yang, ``{MRLoc: Mitigating Row-Hammering Based on Memory
  Locality},'' in \emph{DAC}, 2019.

\bibitem{zhang2019telehammer}
Z.~Zhang \emph{et~al.}, ``{TeleHammer: A Stealthy Cross-Boundary Rowhammer
  Technique},'' arXiv:1912.03076 [cs.CR], 2019.

\bibitem{zhang2020pthammer}
Z.~Zhang \emph{et~al.}, ``{PThammer: Cross-User-Kernel-Boundary Rowhammer
  through Implicit Accesses},'' in \emph{MICRO}, 2020.

\bibitem{zuravleff1997controller}
W.~K. Zuravleff and T.~Robinson, ``{Controller for a Synchronous DRAM That
  Maximizes Throughput by Allowing Memory Requests and Commands to Be Issued
  Out of Order},'' {U.S.}\ Patent 5,630,096. 1997.

\end{thebibliography}
